%% file: multiplicity_pPb_preprint.tex
\newcommand*{\ATLASLATEXPATH}{latex/}
\documentclass[cernpreprint,texlive=2011,txfonts,UKenglish]{latex/atlasdoc}
\pdfoutput=1

\usepackage[compress,comma,numbers]{natbib}
\usepackage[biblatex=false]{\ATLASLATEXPATH atlaspackage}
\usepackage{latex/atlasphysics} 

\graphicspath{{logos/}{figures/}}
\newcommand{\paper}{paper}

\AtlasTitle{Measurement of the centrality dependence of the charged-particle pseudorapidity distribution in proton--lead collisions  at $\sqn = 5.02$~TeV with the ATLAS detector}

\PreprintIdNumber{CERN-PH-EP-2015-160}

\AtlasJournal{EPJC}

\AtlasAbstract{The centrality dependence of the mean charged-particle multiplicity
as a function of pseudorapidity is measured in approximately 1 $\mu$b$^{-1}$ of proton--lead collisions at a
nucleon--nucleon centre-of-mass energy of $\sqn = 5.02$~\TeV\ using the
ATLAS detector at the Large Hadron Collider.  Charged particles with absolute 
pseudorapidity less than 2.7 are reconstructed using the
ATLAS pixel detector. The \pPb\ collision centrality is
characterised by the total transverse energy measured in the Pb-going
direction of the forward calorimeter.  
The charged-particle pseudorapidity distributions are
found to vary strongly with centrality, with an increasing asymmetry between the proton-going
and Pb-going directions as the collisions become more central. Three
different estimations of the number of nucleons participating in the \pPb\ collision have
been carried out using the Glauber model as well as two
Glauber--Gribov inspired extensions to the Glauber model. Charged-particle multiplicities per
participant pair are found to vary differently for these three models, highlighting the importance of
including colour fluctuations in nucleon--nucleon collisions in the modelling of
the initial state of \pPb\ collisions.}

\begin{document}

\maketitle

\tableofcontents


\hyphenation{simi-lar pseudo-ra-pi-di-ty calo-ri-meter}

\section{Introduction}
\input{1_introduction}

\section{Experimental setup}
\label{sec:exp}
\input{2_experiment}

\section{Event selection}
\label{sec:ana}
\input{3_dataset}

\section{Measurement of charged-particle multiplicity}
\label{sec:reco}
\input{4_reconstruction}

\section{Results}
\label{sec:results}
\input{6_results}

\section{Conclusions}
\label{sec:conc}
\input{7_conclusions}

\section*{Acknowledgements}
\input{Acknowledgements2015-09-15}

\clearpage

\appendix
\part*{Appendix: Glauber model analysis}
\addcontentsline{toc}{section}{Appendix}
\input{5_glauber}

\bibliographystyle{bibtex/bst/atlasBibStyleWoTitle}
\bibliography{multiplicity_pPb_paper}

\clearpage 
\input{atlas_authlist}

\end{document}

%% file: 1_introduction.tex
Proton--nucleus (\pA) collisions at the Large Hadron Collider
(LHC)~\cite{Evans:2008zzb} provide an opportunity to probe the physics
of the initial state of ultra-relativistic heavy-ion (A+A) collisions 
without the presence of thermalisation and collective
evolution \cite{Csernai:2006zz}.  In particular, \pA\ measurements can
provide insight into the effect of an extended nuclear target on the
dynamics of soft and hard scattering processes and subsequent particle
production.  Historically, measurements of the average charged-particle 
multiplicity as a function of pseudorapidity, \dndeta, where pseudorapidity 
is defined as $\eta=-\ln\tan(\theta/2)$ with $\theta$ the particle angle 
with respect to the beam direction, have
yielded important insight into soft particle production dynamics in
proton-- and deuteron--nucleus (\pdA) collisions~\cite{Busza:1989px,Elias:1979cp,DeMarzo:1982rh,Brick:1989dm,Back:2003hx, ALICE:2012mj}
and provided essential tests of models of inclusive soft hadron production.

Additional information is obtained if measurements of the charged-particle multiplicities are presented as a function of centrality, an
 experimental quantity that characterises the \pdA\ collision
 geometry. 
 Previous measurements in \dAu\
 collisions at the Relativistic Heavy Ion Collider (RHIC)~\cite{Hahn2003245} have
 characterised the centrality using particle multiplicities at large
 pseudorapidity, either symmetric around
 mid-rapidity~\cite{Back:2004mr} or in the Au fragmentation
 direction~\cite{Adler:2007aa}.  
These measurements have shown that the rapidity-integrated particle 
multiplicity in \dAu\ collisions scales with the number of inelastically 
interacting, or ``participating", nucleons, \Npart. 
  This scaling behaviour has been
 interpreted as the result of coherent multiple soft interactions of
 the projectile nucleon in the target nucleus, and is known as the
 wounded--nucleon (WN) model~\cite{Bialas:1976ed}.  The charged-particle multiplicity distributions as a function of pseudorapidity
 measured in central \dAu\ collisions are asymmetric and peaked in the
 Au-going direction \cite{Back:2003hx}.  This observation has been
 explained using well-known phenomenology of soft hadron
 production~\cite{Adil:2005qn}.

There are alternative descriptions of the centrality dependence of
the \dndeta distribution in \dAu\ collisions at
RHIC~\cite{Tribedy:2011aa,Albacete:2013tpa} and \pPb\ collisions
at the LHC \cite{Albacete:2013tpa,ALICE:2012xs,Rezaeian:2012ye} based on parton
saturation models.  Measurements of the centrality dependence
of \dndeta distributions in \pPb\ collisions provide an essential test of soft
hadron production mechanisms at the LHC. Such tests have become of greater 
importance given the observation of
two-particle~\cite{CMS:2012qk,Abelev:2012ola,Aad:2012gla,Chatrchyan:2013nka}
and multi-particle
\cite{Chatrchyan:2013nka,Abelev:2014mda,Aad:2013fja} correlations 
in the final state of \pPb\ collisions at the LHC. These correlations
are currently interpreted as resulting from either initial-state  
saturation
effects~\cite{Albacete:2013tpa,Dusling:2012wy,Dusling:2013oia} or  
from the collective dynamics of the final
state~\cite{Bozek:2013uha,Shuryak:2013ke,Bzdak:2013zma,Qin:2013bha,Broniowski:2013rka}. For
either interpretation, information on the centrality dependence
of \dndeta can provide important input for determining the mechanism
responsible for these structures. 

Recent measurements from the ALICE experiment~\cite{Adam:2014qja} 
show behaviour in the centrality dependence of the charged-particle pseudorapidity distributions, which is qualitatively similar to that observed at RHIC.
That analysis compared different methods for characterising centrality
and suggested that the method used to define centrality may have a significant 
impact on the centrality dependence of the measured \dndeta distribution.

An important component of any centrality-dependent analysis is the
geometric model used to relate experimental observables to the
geometry of the nuclear collision. Glauber Monte Carlo 
models~\cite{Miller:2007ri}, which simulate the interactions of the
incident nucleons using a semi-classical eikonal approximation, have
been successfully applied to many different \nucnuc measurements at
RHIC and the LHC. A key parameter of such models is the
inelastic nucleon--nucleon cross-section, which is taken to be 70~mb
for this analysis~\cite{Adam:2014qja}.  However, the Glauber
multiple-scattering approximation assumes that the nucleons remain on
the mass shell between successive scatterings, and this assumption is
badly broken in ultra-relativistic collisions. Corrections to the
Glauber model \cite{Gribov:1968jf}, hereafter referred to as
``Glauber--Gribov,'' are needed to account for the off-shell propagation of
the nucleons between collisions. 

A particular implementation of the
Glauber--Gribov approach is provided by the colour-fluctuation model 
\cite{Heiselberg:1991is,Blaettel:1993ah,Guzey:2005tk,Alvioli:2013vk}. That model 
accounts for event-to-event fluctuations in the configuration of 
the incoming proton that are assumed to be frozen over the timescale
of a collision and that can change the effective cross-section with
which the proton scatters off nucleons in the nucleus. These event-by-event
fluctuations in the cross-section can be represented by a probability
distribution $P(\sigma)$. The width of that distribution can be
characterised by a parameter \omsig, which is the relative variance of
the $\sigma$ distribution,
$\omsig \equiv \langle \left(\sigma/\sigma_{\mathrm{tot}} - 1\right)^2 \rangle$.
The usual total cross-section, $\sigma_{\mathrm{tot}}$, is the
event-averaged cross-section, or, equivalently, the first moment of the
$P(\sigma)$ distribution, $\sigma_{\mathrm{tot}}
= \int_0^{\infty}{d\sigma} P(\sigma)\, \sigma$. 
The parameter \omsig can be measured using diffractive
proton--proton scattering at high energy \cite{Blaettel:1993ah,Guzey:2005tk}. 
First estimates of \omsig\ at LHC energies \cite{Guzey:2005tk}
extrapolated to 5~TeV yielded $\omsig \sim 0.11$, while a more recent analysis suggested
$\omsig \sim 0.2$ \cite{Alvioli:2013vk}. Because the cross-section
fluctuations in the Glauber--Gribov colour-fluctuation (GGCF) model may
have a significant impact on the interpretation of the results of this
analysis, the geometry of \pPb\ collisions has been
evaluated using both the standard Glauber model and the GGCF model 
with $\omsig = 0.11$ and 0.2. 

This \paper\ presents measurements of the centrality dependence
 of \dndeta in \pPb\ collisions at $\sqn = 5.02$~\TeV\ using
 $1~\mu{\rm b}^{-1}$ of data recorded by the ATLAS experiment~\cite{Aad:2008zzm} in
 September 2012. Charged particles are detected in the ATLAS pixel
 detector and are reconstructed using a two-point tracklet algorithm
 similar to that used for the \PbPb\ multiplicity
 measurement~\cite{ATLAS:2011ag}.  Measurements of \dndeta are
 presented for several intervals in collision centrality characterised
 by the total transverse energy measured in the forward section of the
 ATLAS calorimeter on the Pb-going side of the detector. A standard
 Glauber model~\cite{Miller:2007ri} and the GGCF
 model~\cite{Guzey:2005tk,Alvioli:2013vk} with $\omsig = 0.11$ and 0.2
 are used to estimate \avgNpart\ for each centrality interval,
 allowing a measurement of the \Npart\ dependence of the charged-particle multiplicity.

The \paper\ is organised as follows. 
Section~\ref{sec:exp} describes the subdetectors of the ATLAS experiment relevant for this measurement. 
Section~\ref{sec:ana} describes the event selection. 
Section~\ref{sec:mc_set} describes the Monte Carlo simulations used to understand the performance and derive the corrections to the measured quantities. 
Section~\ref{sec:centrality} describes the choice of centrality variable. 
Section~\ref{sec:reco} describes the measurement of the charged-particle multiplicity and Sect.~\ref{sec:syst} describes the estimation of the systematic uncertainties. 
Section~\ref{sec:results} presents the results of the measurement, and the interpretation of the yields of charged particles per participant is discussed in Sect.~\ref{sec:perNpartyields}. 
Section~\ref{sec:conc} concludes the \paper. 
 The estimation of the geometric parameters in each centrality 
 interval for the Glauber and GGCF models is presented in detail 
 in the Appendix.

%% file: 2_experiment.tex
The ATLAS detector is described in detail in Ref.~\cite{Aad:2008zzm}. 
The data selection and analysis presented in this \paper\ is performed using the
ATLAS inner detector (ID), calorimeters, minimum-bias trigger
scintillators (MBTS), and the trigger system. 
The inner detector measures
charged-particle tracks using a combination of silicon pixel detectors,
silicon microstrip detectors (SCT), and a straw-tube
transition-radiation tracker (TRT), all immersed in a 2~T axial magnetic
field.  The pixel detector 
is divided into ``barrel'' and ``endcap'' sections.  For collisions occurring at the
nominal interaction point,\footnote{ATLAS uses a right-handed coordinate system with its origin
at the nominal interaction point (IP) in the centre of the detector
and the $z$-axis along the beam pipe. The $x$-axis points from the IP
to the centre of the LHC ring, and the $y$-axis points
upward. Cylindrical coordinates $(r,\phi)$ are used in the transverse
plane, $\phi$ being the azimuthal angle around the beam pipe. The
pseudorapidity is defined in terms of the polar angle $\theta$ as
$\eta=-\ln\tan(\theta/2)$. 
} 
 the barrel section of the pixel detector
allows measurements of charged-particle tracks over $|\eta| < 2.2$.
 The endcap sections extend the detector coverage, spanning the
pseudorapidity interval $1.6<|\eta|<2.7$.  The SCT and TRT detectors cover
$|\eta| < 2.5$ and $|\eta| < 2$, respectively, also through a combination
of barrel and endcap sections. 

The barrel section of the pixel detector consists of
three layers of staves at radii of $50.5$~mm, $88.5$~mm, and $122.5$~mm from the
nominal beam axis, and extending $\pm 400.5$~mm from the
centre of the detector in the $z$ direction.  The endcap consists of
three disks placed symmetrically on each side of the interaction
region at $z$ locations of $\pm 493$~mm, $\pm 578$~mm and $\pm 648$~mm from
the centre of the detector.  All pixel sensors in the
pixel detector, in both the barrel and endcap regions, are identical and
have a nominal size of $50\,{\rm \mu m} \times400\,{\rm \mu m}$.

The MBTS detect charged particles in the range $2.1 < |\eta| < 3.9$
using two hodoscopes, each of which is subdivided into 16 counters
positioned at $z=\pm3.6$~m.  The ATLAS calorimeters cover the full
azimuth and the pseudorapidity range $|\eta|<4.9$ with the forward 
part (FCal) consisting of two modules positioned on either side of the 
interaction region and covering $\FCalLowEta < |\eta| < 4.9$.  The FCal
modules are composed of tungsten and copper absorbers with liquid
argon as the active medium, which together provide 10 interaction
lengths of material.

The LHC delivered its first proton--nucleus collisions in a short \pPb\
``pilot'' run at $\sqn = 5.02$~\TeV\ in September 2012. During that
run the LHC was configured with a clockwise 4~\TeV\ proton beam and an
anti-clockwise 1.57~\TeV\ per-nucleon Pb beam that together produced
collisions with a nucleon--nucleon centre-of-mass energy of $\sqn =
5.02$~\TeV\ and a longitudinal rapidity boost of 0.465 units with
respect to the ATLAS laboratory frame. Following a common convention
used for \pA\ measurements, the rapidity is taken to be positive in the
direction of the proton beam, i.e. opposite to the usual ATLAS
convention for \pp\ collisions. With this convention, the ATLAS laboratory 
frame rapidity $y$ and the \pPb\ centre-of-mass system rapidity 
$y_{\text{cm}}$ are related as $y_{\text{cm}} = y-0.465$.

%% file: 3_dataset.tex
Minimum-bias \pPb\ collisions were selected by a trigger that required
a signal in at least two MBTS counters. The \pPb\ events selected for
analysis are required to have at least one hit in each side of the
MBTS, a difference between the times measured
in the two MBTS hodoscopes of less than 10~ns, and a reconstructed
collision vertex in longitudinal direction, \zvtx, within 175~mm of the nominal centre of 
the ATLAS detector. Collision vertices are defined using charged-particle
tracks reconstructed by an algorithm optimised for \pp\
minimum-bias measurements~\cite{Aad:2010ac}. Reconstructed vertices 
are required to have at least two tracks with transverse momentum 
$p_\mathrm{T} > 0.4$~\GeV. Events containing
multiple \pPb\ collisions are rare due to very low instantaneous luminosity 
during the pilot run and are further suppressed in the analysis by rejecting
events with two collision vertices that are separated in $z$ by more
than 15~mm. Applying this selection reduces the fraction of events with 
multiple collisions from less than 0.07\% to below 0.01\%.

To remove potentially significant contributions from electromagnetic
and diffractive processes, the topology of the events was first
analysed in a manner similar to that performed in a measurement of
rapidity gap cross-sections in 7~\TeV\ proton--proton
collisions~\cite{Aad:2012pw}. The pseudorapidity coverage of the
calorimeter, $-4.9 < \eta < 4.9$, is divided into $\Delta \eta = 0.2$
intervals, and each interval containing one or more clusters with
\pt greater than 0.2~\GeV\ is considered as
occupied. To suppress the contributions from noise, clusters are 
considered only if they contained at least one cell with an energy at
least four times the standard deviation of the cell noise
distribution.

Then, the edge-gap on the
Pb-going side of the detector is calculated as the distance in
pseudorapidity between the detector edge $\eta=-4.9$ and the nearest
occupied interval.
Events with edge-gaps larger than two units of pseudorapidity
typically result from electromagnetic or diffractive excitation of the
proton and are removed from the analysis. The effect of this selection
is identical to the requirement of a cluster with transverse energy $\ET>0.2$~\GeV\ 
to be present in the region $\eta< -2.9$.
No requirement is imposed on edge-gaps on the proton-going side. The gap requirement removes, with
good efficiency, a sample of events which are not naturally described
in a Glauber picture of \pPb collisions. 
This requirement removes a further fraction $\fgap = 1$\% of the events passing the 
vertex and MBTS cuts, yielding a total of 2.1~million events for this analysis. 
The result of this
event selection is to isolate a fiducial class of \pPb events, defined
as inelastic \pPb events that have a suppressed contribution from 
diffractive proton excitation events.

\section{Monte Carlo simulation}
\label{sec:mc_set}
The response of the ATLAS detector and the performance of
the charged-particle reconstruction algorithms are evaluated using one million minimum-bias
5.02~\TeV\ Monte Carlo (MC) \pPb\ events, produced by version 1.38b of
the {\sc Hijing} event generator \cite{Wang:1991hta} with diffractive
processes disabled. The four-momentum of each generated particle is
longitudinally boosted by a rapidity of 0.465 to match the beam
conditions in the data.  The detector response to these events is fully simulated using
{\sc Geant4}~\cite{Agostinelli:2002hh,Aad:2010ah}. The resulting events are
digitised using conditions appropriate for the pilot \pPb\ run and
fully reconstructed using the same algorithms that are applied to the
experimental data. This MC sample is primarily used to evaluate the
efficiency of the ATLAS detector for the charged-particle measurements.

The detector response and event selection efficiencies for peripheral and 
diffractive \pPb\ events have properties similar to those for inelastic or 
diffractive \pp\ collisions, respectively. To evaluate these responses and 
efficiencies, the \pp\ samples are generated at $\sqs = 5.02$~\TeV\ with 
particle kinematics boosted to match the \pPb\ beam conditions. 
Separate samples of 
minimum-bias, single-diffractive, and double-diffractive \pp\ collisions
with one million events each are produced using both {\sc Pythia6} \cite{Sjostrand:2006za} (version 6.425,
AMBT2 parameter set (tune)~\cite{ATL-PHYS-PUB-2011-009}, 
CTEQ6L1 PDF~\cite{Pumplin:2002vw}) and {\sc Pythia8} \cite{Sjostrand:2007gs} 
(version 8.150, 4C tune~\cite{ATL-PHYS-PUB-2012-003}, 
MSTW2008LO PDF~\cite{Martin:2009iq}), and simulated, digitised and reconstructed in the same 
manner as the \pPb\ events. These six samples are primarily used for
the Glauber model analysis described in the Appendix.

\section{Centrality selection}
\label{sec:centrality}
\label{sec:centrality_def}

\begin{figure}[!htb]
\centerline{
\includegraphics[width=0.60\textwidth]{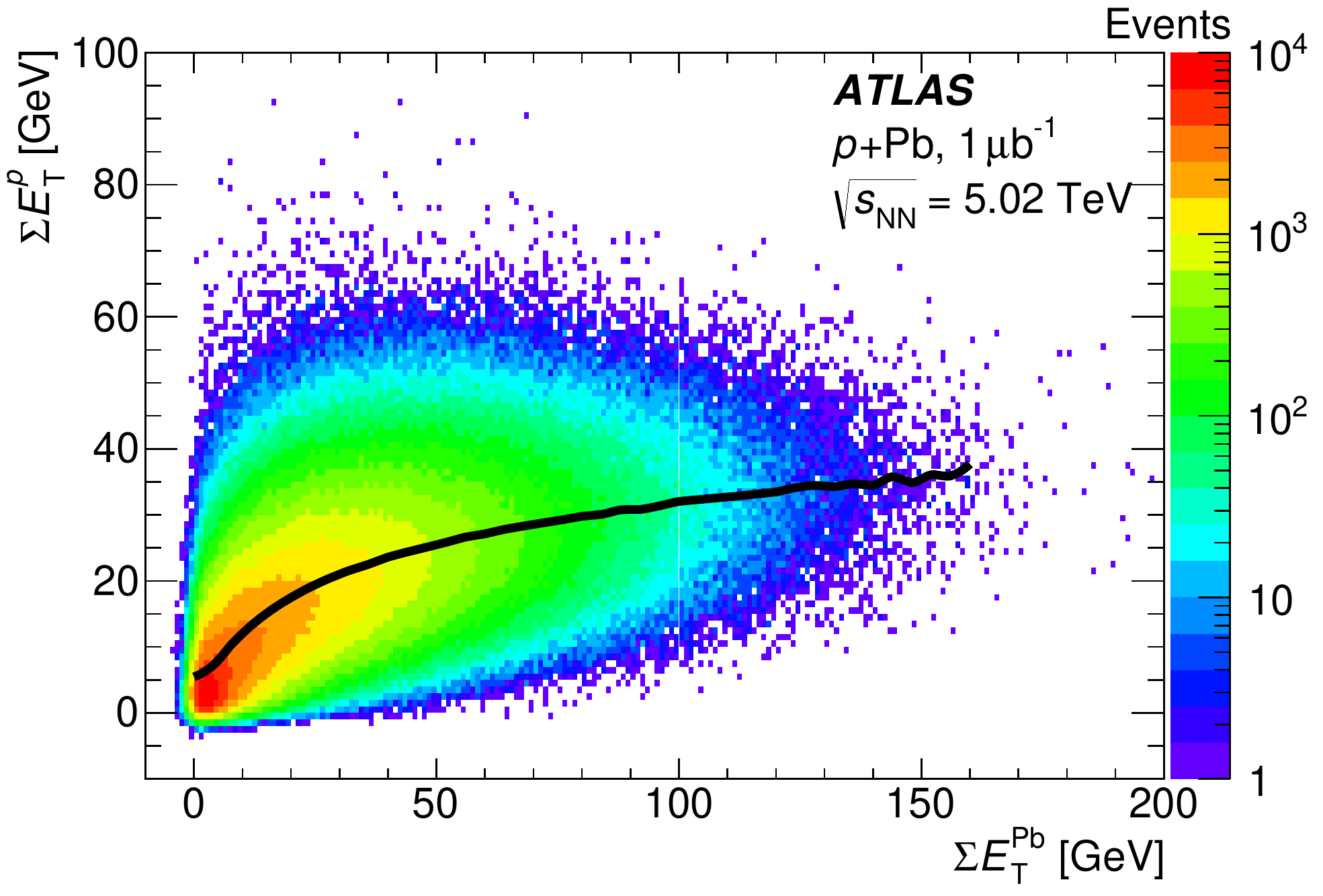}
}
\caption{Distribution of proton-going (\sumETp) versus Pb-going (\sumETPb) total transverse energy in the forward calorimeter for \pPb\ collisions included in this analysis. The curve shows the average \sumETp\ as a function of \sumETPb.}
\label{fig:fcalCvsA}
\end{figure}

For \PbPb\ collisions, the ATLAS experiment uses the total transverse energy, \sumET,
measured in the two forward calorimeter sections to characterise the collision 
centrality \cite{ATLAS:2011ah}.  However, the intrinsic asymmetry of
the \pPb\ collisions and the rapidity shift of the centre-of-mass
causes an asymmetry in the soft particle production measured on the
two sides of the calorimeter. Figure~\ref{fig:fcalCvsA}
shows the correlation between the summed transverse energies measured in the
proton-going ($3.1 < \eta < 4.9 $) and Pb-going ($-4.9 < \eta < -3.1$)
directions, \sumETp\ and \sumETPb, respectively. The transverse energies 
are evaluated at an energy scale calibrated for electromagnetic showers and have not been 
corrected for hadronic response. 
Figure~\ref{fig:fcalCvsA} shows that the mean \sumETp\ rapidly flattens with increasing 
\sumETPb\ for $\sumETPb \gtrsim 30$~\GeV, 
indicating that \sumETp\ is less sensitive than
\sumETPb\ to the increased particle production expected to result from
multiple interactions of the proton in the target nucleus in central
collisions. 
Thus, \sumETPb\ alone, rather than $\sumETPb+\sumETp$, is chosen as the primary quantity used to characterise \pPb\ collision centrality for the measurement presented in this paper. However, we describe alternate choices of the centrality-defining region below and 
evaluate the sensitivity of the measurement to this definition.

\begin{figure}[!ht]
\centerline{
\includegraphics[width=0.60\textwidth]{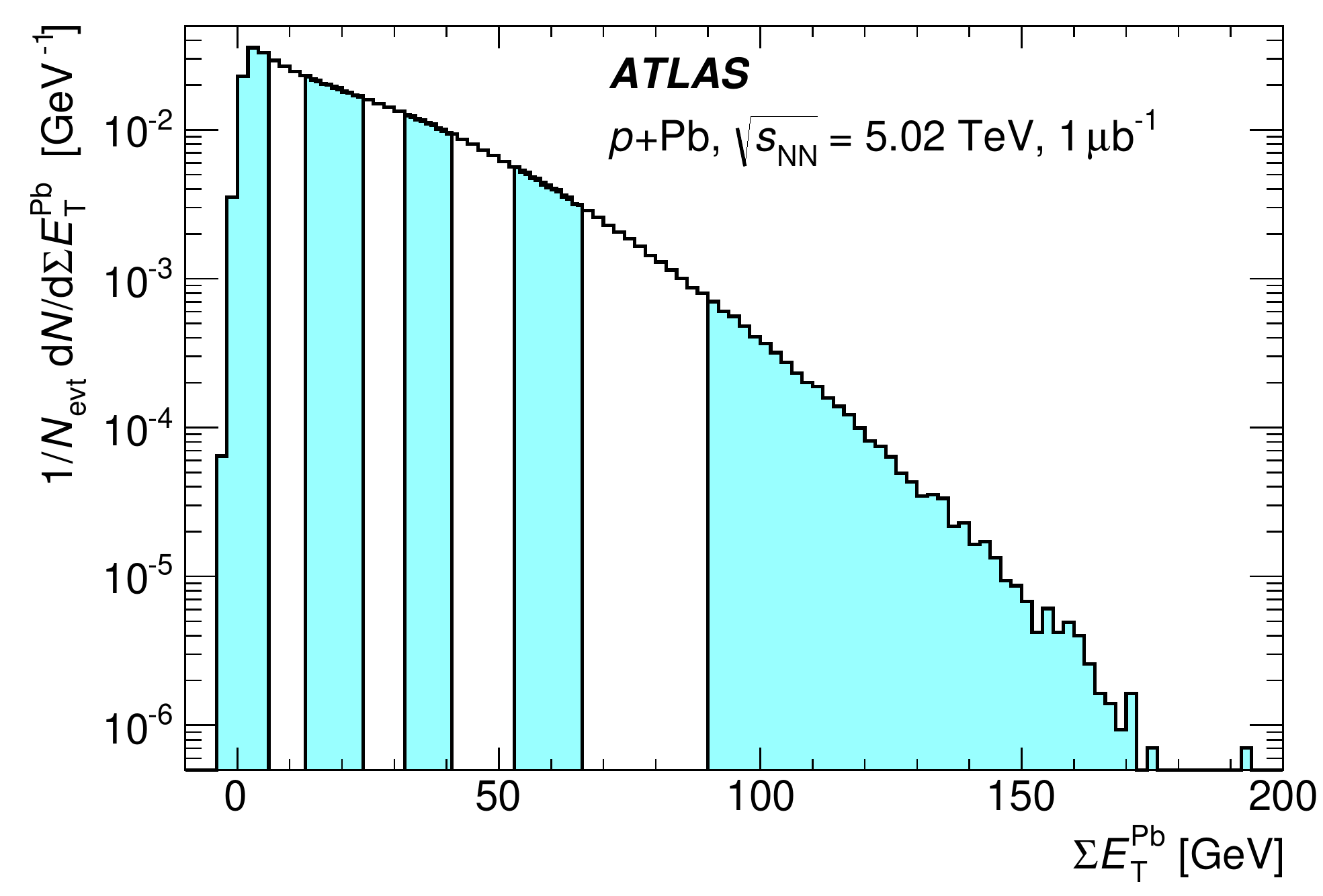}
}
\caption{Distribution of the Pb-going total transverse energy in the forward calorimeter \sumETPb\ values for events satisfying all analysis cuts including the Pb-going rapidity gap exclusion. The alternating shaded and unshaded bands indicate centrality intervals, from right (central) to left (peripheral), 0--1\%, 1--5\%, 5--10\%, 10--20\%, 20--30\%, 30--40\%, 40--60\%, 60--90\% and the interval 90--100\% that is not used in this analysis.}
\label{fig:dndet}
\end{figure}

The distribution of \sumETPb\ for events passing the \pPb\
analysis selection is shown in Fig.~\ref{fig:dndet}. 
The following centrality intervals are defined in terms of
percentiles of the \sumETPb\ distribution: 0--1\%, 1--5\%,
5--10\%, 10--20\%, 20--30\%, 30--40\%, 40--60\%, 
and 60--90\%. The \sumETPb\ ranges corresponding to these centrality
intervals are indicated by the alternating filled and unfilled
regions in Fig.~\ref{fig:dndet}, with the 0--1\% interval, containing the most central collisions, being
rightmost. Since the composition of the events in the most peripheral
90--100\% interval is not well constrained, these events are excluded from
the analysis. 
The nominal centrality intervals were defined after accounting for a 2\% inefficiency, 
as described in the Appendix, 
for the fiducial class of \pPb events defined above to pass 
the applied event selection. Alternate intervals were also defined by varying this 
estimated inefficiency to 0\% and 4\%, and is used as a systematic check on the results. 
While the inefficiency is confined to the 90--100\%
interval, it influences the \sumETPb ranges associated with each centrality interval.
Potential hard scattering contributions to \sumETPb\ have been evaluated
in a separate analysis~\cite{ATLAS:2014cpa} by explicitly subtracting the contributions
from reconstructed jets that fall partly or completely in the Pb-going
FCal acceptance. That analysis showed negligible impact from hard
scattering processes on the measured \sumETPb\ distribution.

To test the sensitivity of the results 
to the choice
of pseudorapidity interval used for the \sumET\ measurement, two
alternative \sumET\ quantities are defined. The
former, \sumETPbfar, is defined as the total transverse energy in FCal
cells with $\eta < -4.0$. 
The latter,  
\sumETsymm, is defined as the total transverse energy in the two intervals 
$4.0 < \eta < 4.9$ and $-4.0< \eta < -3.1$, 
an approximately symmetric interval 
when expressed in pseudorapidity in the centre of mass system $\eta_{\rm{cm}}$. 
The first of these alternatives is used to evaluate the
potential auto-correlation between the measured charged-particle
multiplicities and the centrality observable by increasing the
rapidity gap between the two measurements. The second is used to
evaluate the differences between
an asymmetric (Pb-going) and symmetric (both sides) centrality observable. 
The effect of these alternative definitions is discussed in Sect.~\ref{sec:results}.

\begin{figure}[!htb]
\centerline{
\includegraphics[width=0.60\linewidth]{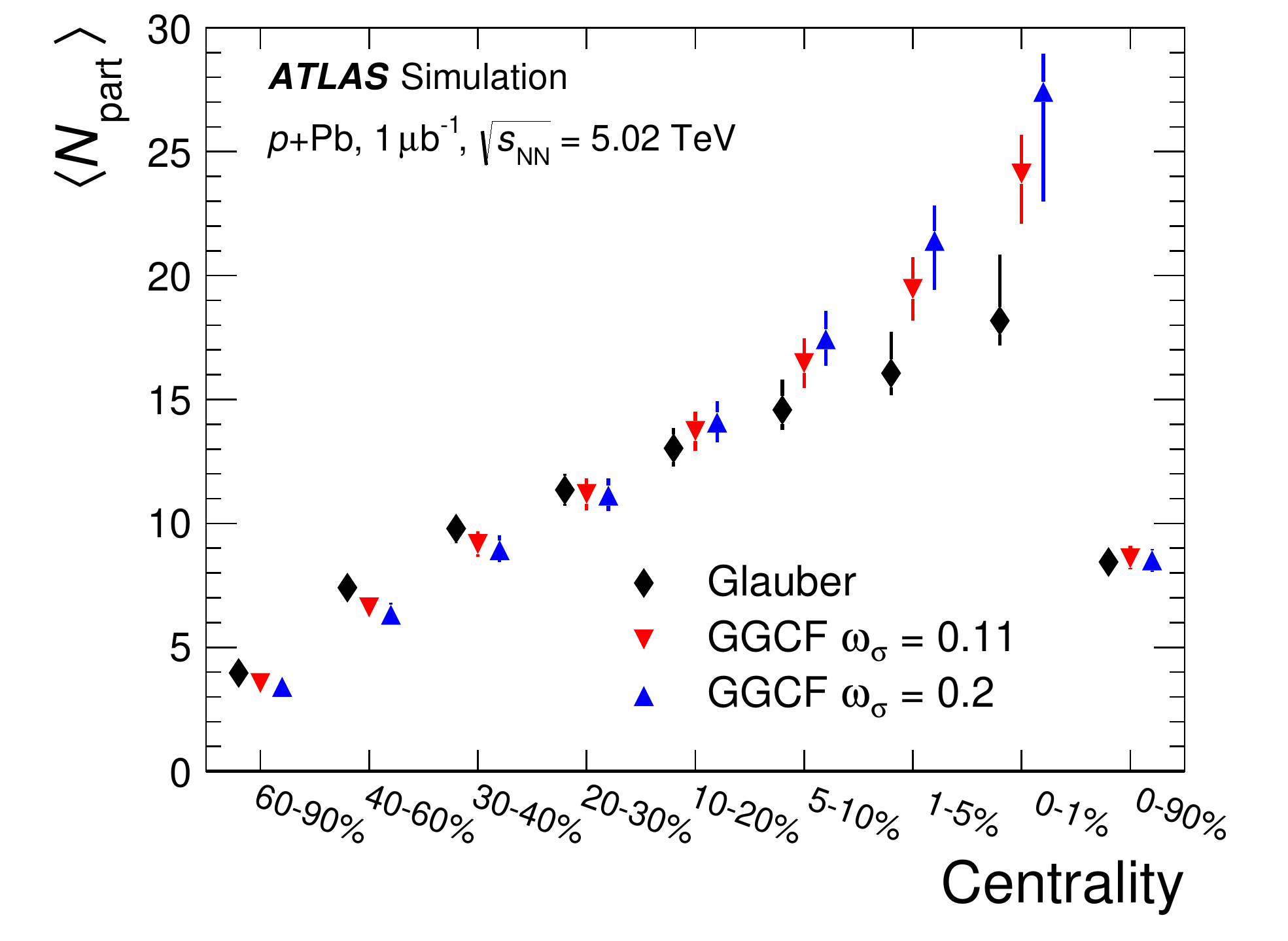}
}
\caption{Mean value of the number of participating nucleons \avgNpart\ for different centrality bins, resulting from fits to the measured \sumETPb\ distribution using Glauber and Glauber--Gribov \Npart\ distributions. The error bars indicate asymmetric systematic uncertainties.}
\label{fig:fitnpart}
\end{figure}

The Glauber analysis~\cite{Miller:2007ri}
was applied to estimate \avgNpart\ for each of the centrality
intervals used in this analysis. A detailed description is given 
in the Appendix; only a brief summary of the
method is given here. The PHOBOS MC program \cite{Alver:2008aq} 
was used to simulate the geometry of
inelastic \pPb\ collisions using both the standard Glauber and GGCF
models. The resulting \Npart\ distributions are convolved with a
model of the \Npart-dependent \sumETPb\ distributions, the parameters of which are
obtained by fitting the measured \sumETPb\
distribution. The average \Npart\
associated with each centrality interval is obtained with systematic
uncertainties. 
The results are shown in Fig.~\ref{fig:fitnpart} for the
Glauber model and for the GGCF model with $\omsig = 0.11$ and 0.2.

%% file: 4_reconstruction.tex
\subsection{Two-point tracklet and pixel track methods}
The measurement of the charged-particle multiplicity
 is performed using only the pixel detector to
maximise the efficiency for reconstructing charged particles with low transverse momenta.
  Two approaches are used in this analysis.  The
first is the two-point tracklet method commonly used in heavy-ion 
collision experiments~\cite{ATLAS:2011ag,Adcox:2000sp,Alver:2010ck}.
Two variants of this method are implemented in this analysis to
construct the \dndeta distribution and to estimate the systematic
uncertainties, as described below.  The second method uses ``pixel
tracks'', obtained by applying the full track reconstruction
algorithm~\cite{Aad:2010bx} only to the pixel detector. The pixel tracking is less efficient
than the tracklet method as is justified later in the text, but provides measurements of the
particle \pT.  The \dndeta distribution measured using pixel tracks provides a
cross-check on the primary measurement that is performed 
using the two-point tracklets.

In the two-point tracklet algorithm, the event vertex and clusters~\cite{Aad:2014yva} on
an inner pixel layer define a search region for clusters in the outer
layers. The algorithm uses all clusters, except the clusters which  
have low energy deposits inconsistent with minimum-ionising particles
originating from the primary vertex. The algorithm also rejects
duplicate clusters resulting from the overlap of the pixel sensors or
arising from a small set of pixels at the centre of the pixel modules
that share readout channels~\cite{Aad:2008zz}. Two clusters in a given
layer of the pixel detector are considered as one if they have an
angular separation 
$\sqrt{(\delta \phi)^2 + (\delta \eta)^2} < 0.02$, 
because they likely result from the passage of a single particle. 

The pseudorapidity and azimuthal angle of the cluster in the innermost
layer ($\eta,\phi$) and their differences between
the outer and inner layers ($\Delta\eta,\Delta\phi$) are taken as the
parameters of the reconstructed tracklet.  
The $\Delta \eta$ of a tracklet is largely determined by the multiple
scattering of the incident particles in the material of the beam pipe
and detector. This effect plays a less significant role in the
$\Delta \phi$ of a tracklet,  which is driven primarily by the bending
of charged particles in the magnetic field, and hence one expects
$\Delta \phi$ to be larger. 
The tracklet selection cuts are:
\begin{eqnarray}
\label{tracklet_criteria}
|\Delta \eta| & < & 0.015, \hspace{2mm} |\Delta \phi|  \hspace{1.5mm}< \hspace{1.5mm}0.1, \label{eq:trackletcuts}\\
|\Delta \eta| & < & |\Delta \phi|. \label{eq:trackletcutmom}
\end{eqnarray}
Keeping tracklets with $|\Delta\phi| < 0.1$ corresponds to accepting
particles with $\pt \gtrsim 0.1$~\GeV. The selection in
Eq.~(\ref{eq:trackletcutmom}) accounts for the momentum dependence of
charged-particle multiple scattering.

The Monte Carlo simulation for the \dndeta analysis is based on
the {\sc Hijing} event generator, which is described in Sect.~\ref{sec:mc_set}. 
The {\sc Hijing} event generator is known to not accurately reproduce the 
measured particle \pT\ distributions. This is addressed by reweighting 
the {\sc Hijing} \pT\ distribution using the ratio of reconstructed spectra
measured with the pixel track method in the data and in the MC simulation. 
The reweighting function is extrapolated below $\pT=0.1$~\GeV\ and applied 
to all generated particles and their decay products. This is done in intervals of centrality and pseudorapidity. Generator-level primary particles are defined as particles with a mean lifetime
$\tau > 0.3\cdot10^{-10}$~s either directly produced in \pPb interactions 
or from subsequent decays of particles with a shorter lifetime. 
This definition is the same as used in previous measurements of charged-particle production in \pp~\cite{Aad:2010ac} 
and \PbPb~\cite{Aad:2015wga} collisions by ATLAS. 
All other particles are defined as secondaries. Tracklets
are classified as primary or secondary depending on whether the
associated generator-level charged particle is primary or secondary. 
Association between the tracklets and the generator-level particles is based on 
the {\sc Geant4} information about hits produced by these particles.
Tracklets that are formed from the random association of hits 
produced by unrelated particles, or hits in the detector which are 
not matched to any generated particle are referred to as ``fake'' tracklets.   

The contribution of fake tracklets is relatively difficult to model in
the simulation, because of the {\it a priori} unknown contributions of
multiple sources, such as noisy clusters or very low energy particles.
To address this problem, the tracklet algorithm is used in two
different implementations referred to as ``\Mo'' and ``\Mt''.  In \Mo,
at most one tracklet is reconstructed for each cluster on the first
pixel layer.  If multiple clusters on the second pixel layer fall
within the search region, the resulting tracklets are merged into a
single tracklet.  This approach reduces, but does not eliminate, the
contribution of fake tracklets that are then accounted for using an 
MC-based correction.
\Mt\ reconstructs tracklets for all combinations of clusters in only two 
pixel layers, the innermost and the next-to-innermost detector layers. 
To account for the fake tracklets arising from random combinations of
clusters, the same analysis is performed after inverting the $x$ and
$y$ positions of all clusters on the second layer with respect to the
primary vertex
$(x-x_{\text{vtx}},y-y_{\text{vtx}})\rightarrow(-(x-x_{\text{vtx}}),-(y-y_{\text{vtx}}))$. The
tracklet yield from this ``flipped'' analysis,
$\Ntptrt^{\mathrm{fl}}$, is then subtracted from the original tracklet
yield, $\Ntptrt^{\mathrm{ev}}$ to obtain an estimated yield of true 
tracklets \Ntptr,
\begin{equation}
\Ntptr(\eta) = \Ntptrt^{\mathrm{ev}}(\eta) - \Ntptrt^{\mathrm{fl}} (\eta).
\label{eq:flipsub}
\end{equation}

\begin{figure*}[!htb]
\begin{center}
\includegraphics[width=0.49\textwidth]{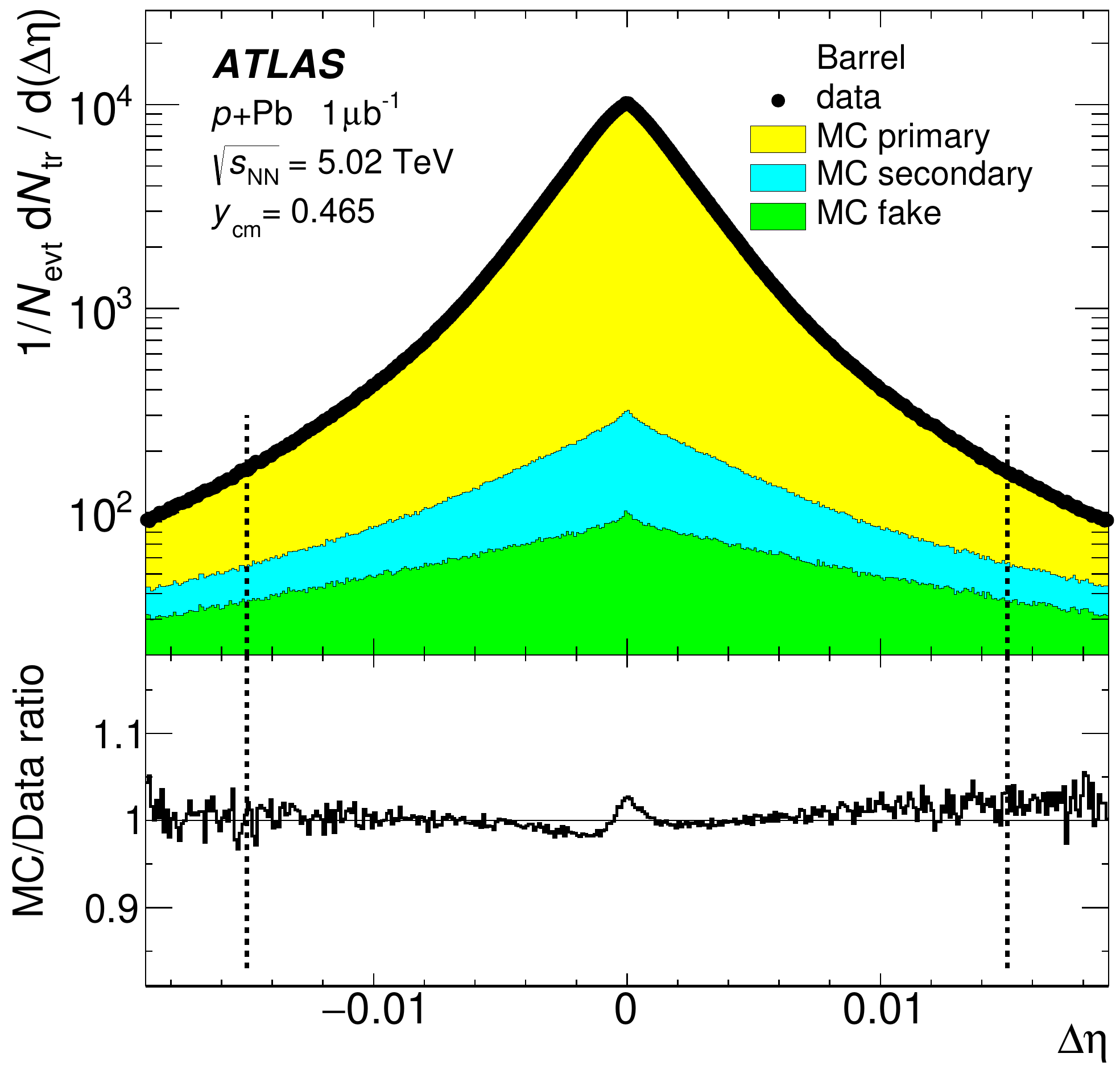}
\includegraphics[width=0.49\textwidth]{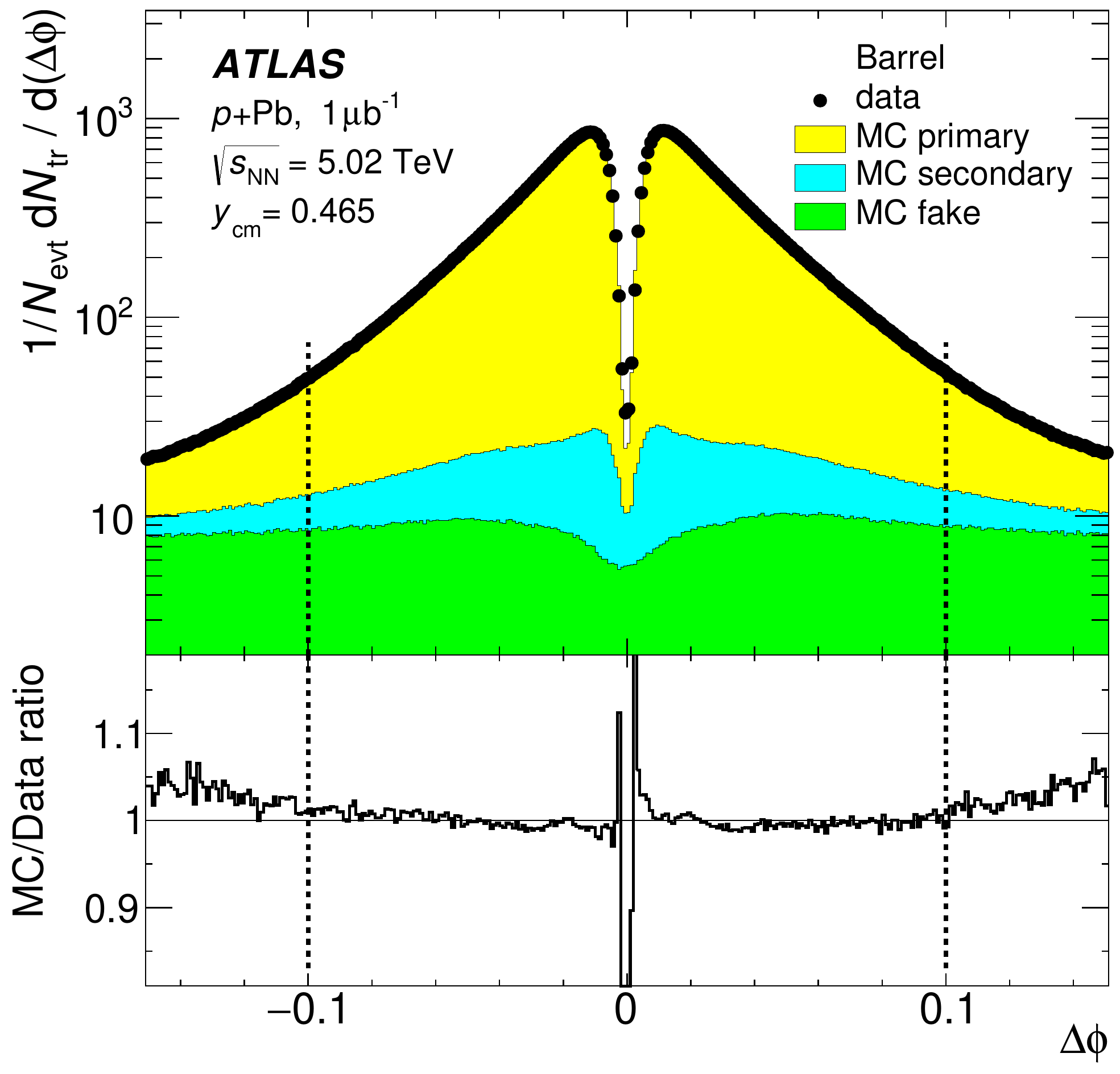}
\includegraphics[width=0.49\textwidth]{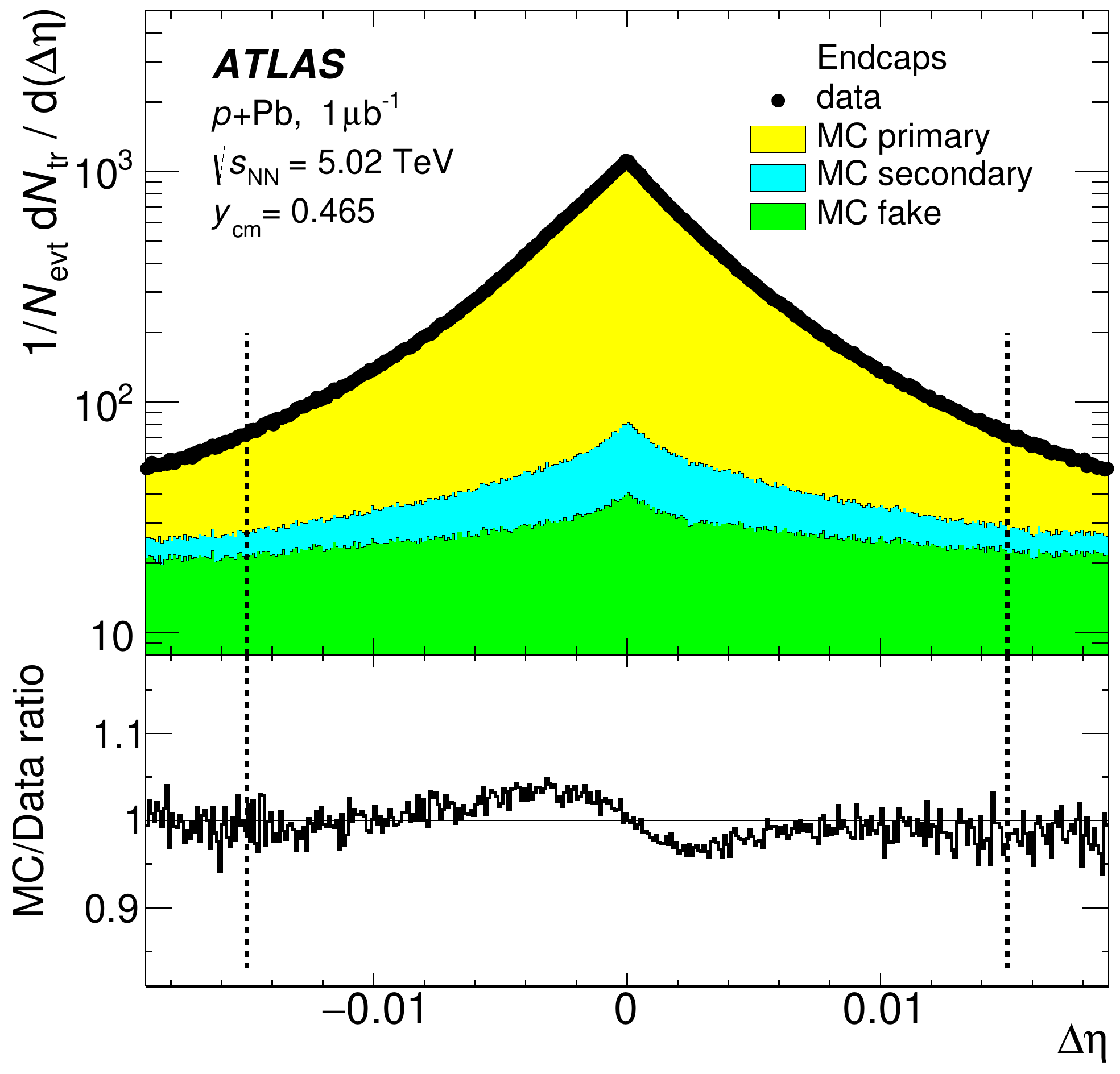}
\includegraphics[width=0.49\textwidth]{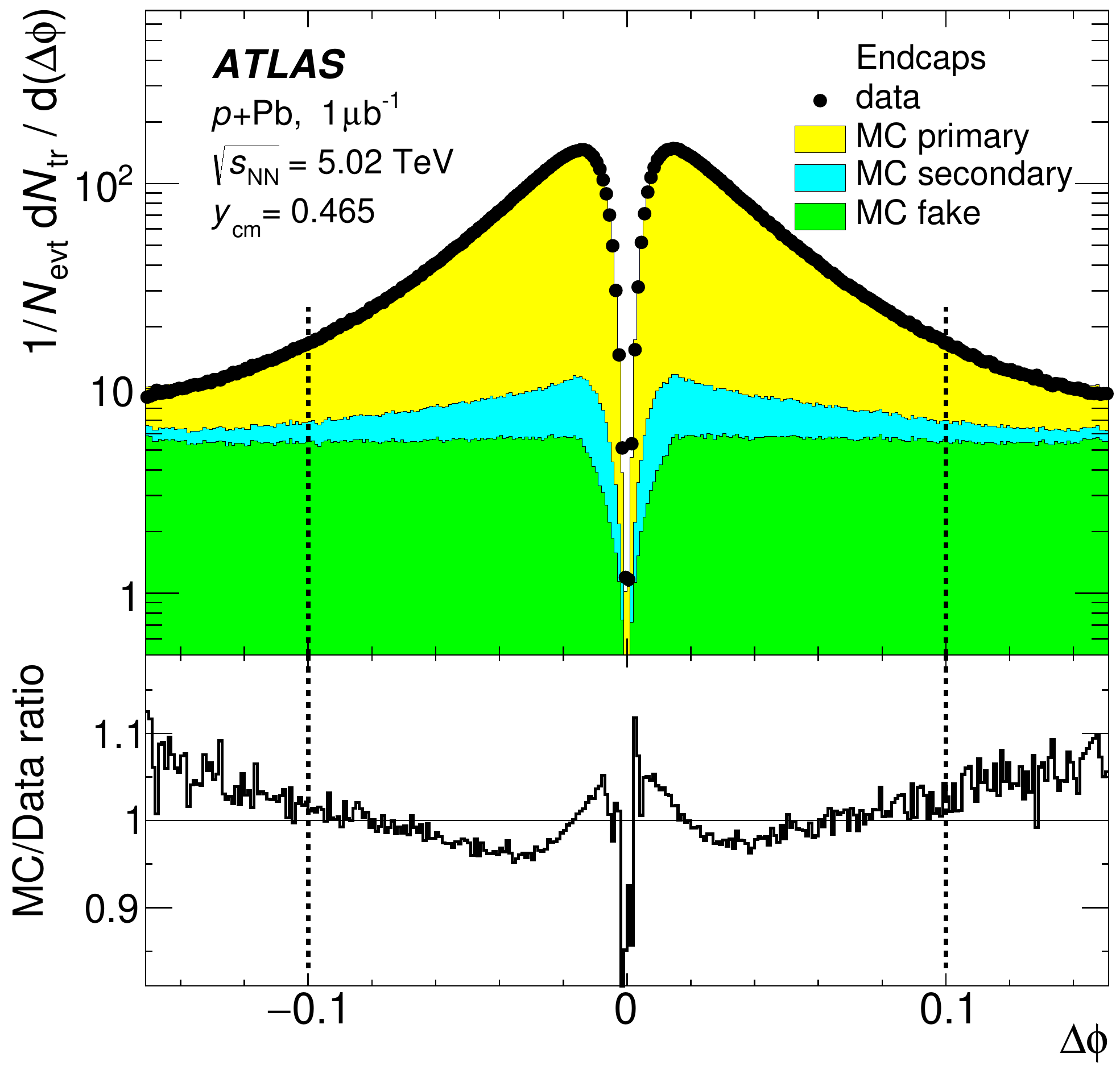}
\caption{Stacked histograms for the differences between the hits of the tracklet in outer and inner detector layers in pseudorapidity $\Delta\eta$ (left) and in azimuth $\Delta\phi$ (right) for the tracklets reconstructed with \Mo\ measured in the data (points) and simulation (histograms) in \pPb\ collisions at \sqn=5.02\,\TeV\ for barrel (top) and endcaps (bottom). Contributions from primary, secondary, and fake tracklets in the simulation are shown separately. The lower panels show the ratio of the simulation to the data.}
\label{fig:deta_phi_mc_data}
\end{center}
\end{figure*}

Distributions of $\Delta \eta$ and $\Delta \phi$ of reconstructed tracklets using 
\Mo\ for data and simulated events are shown in Fig.~\ref{fig:deta_phi_mc_data} 
for the barrel (upper plots) and endcap (lower plots) parts of the pixel detector.
The simulation results show the three contributions from primary,
secondary and fake tracklets. The selection
criteria specified by Eq.~(\ref{eq:trackletcuts}) are shown in
Fig.~\ref{fig:deta_phi_mc_data} as vertical lines and applied in
$\Delta\phi$ for $\Delta\eta$ plots and vice versa. Outside those
lines, the contributions from secondary and fake tracklets are more
difficult to take into account, especially in the endcap region. 
These contributions partially arise from low-\pT\ particles on spiral
trajectories and their description in the MC simulation is therefore very sensitive 
to the amount of detector material. The ratio between simulation and the 
data is also shown for each plot. These ratios are closer to unity in the barrel 
region than in the endcap region, where they deviate by up to 5\% 
except at very low $|\Delta\phi|$. At low $|\Delta\phi|$ corresponding to high \pT, 
the MC deviates from the data even after reweighing procedure based on pixel tracks. 
This is due to low resolution of pixel track at high \pT, however, 
the contribution of high-\pT\ particles to \dndeta\ is negligible.

\begin{figure*}[!htb]
\begin{center}
\includegraphics[width=0.49\textwidth]{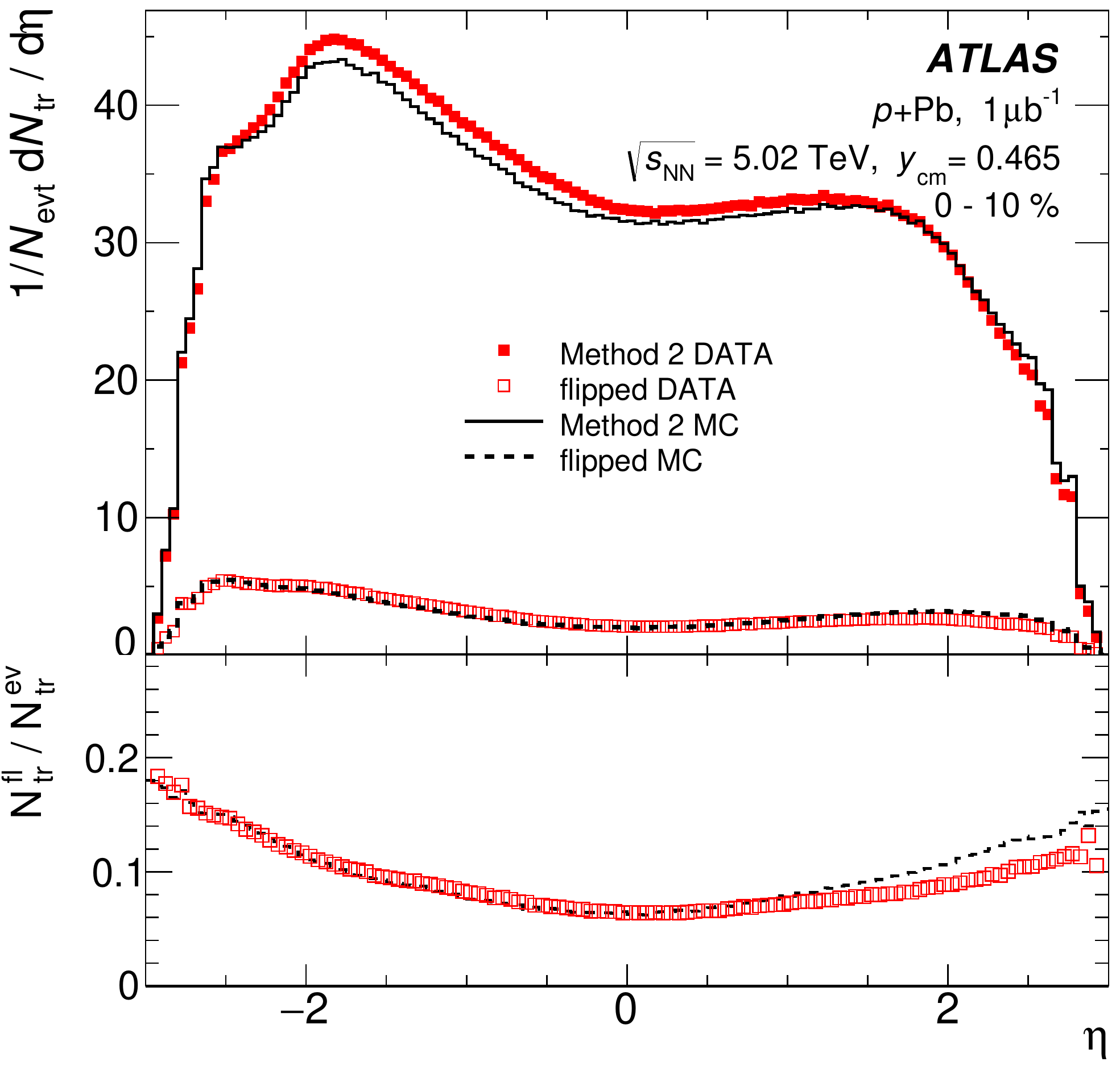}
\includegraphics[width=0.49\textwidth]{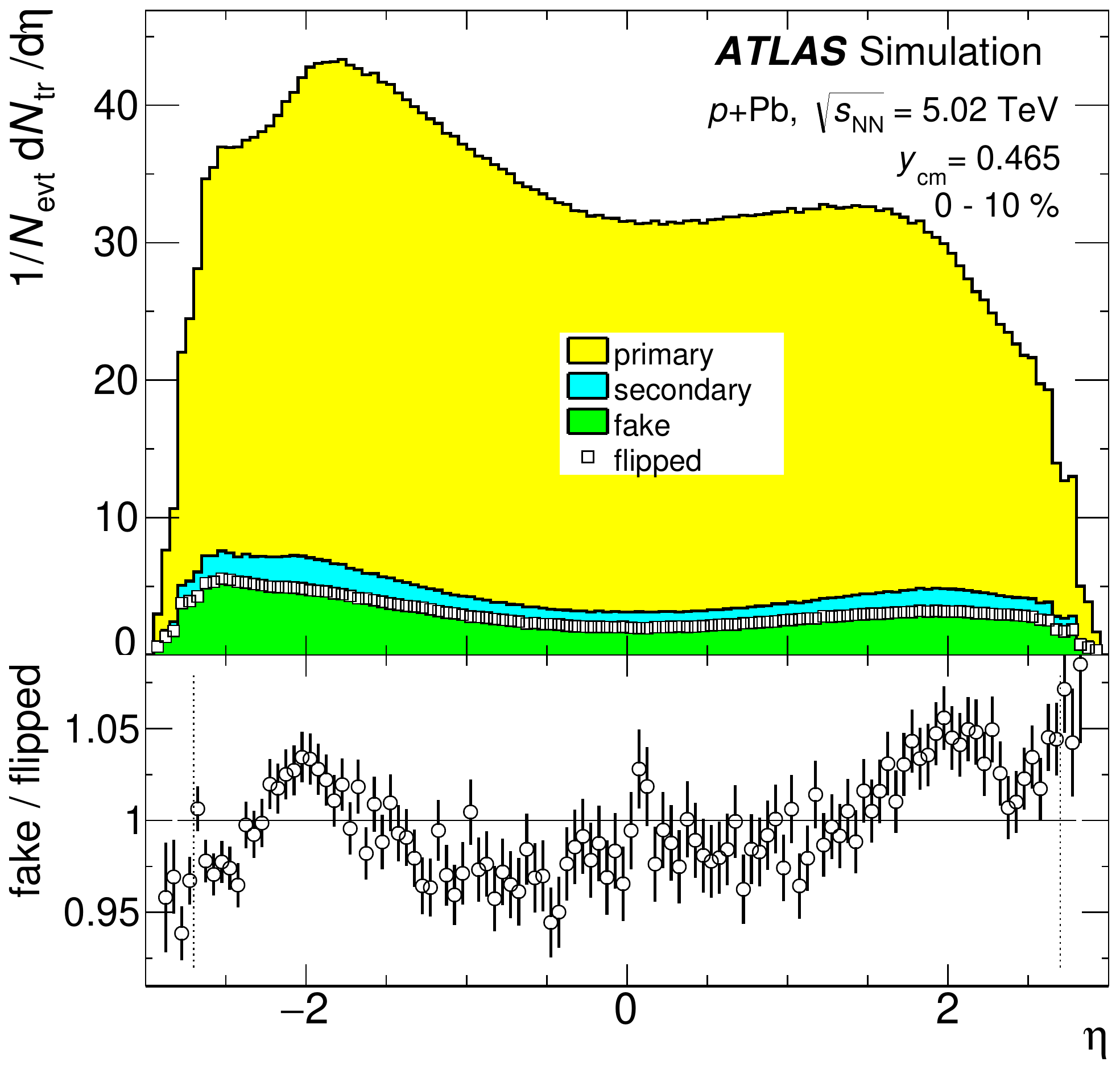}
\caption{$\eta$-distribution of the number of tracklets reconstructed with \Mt. Left top panel: comparison of the simulation (lines) to the data (markers). The results of the flipped reconstruction are shown with open markers for data and dashed line for simulation. Right top panel: the simulated result for three contributions: primary, secondary and fake tracklets. Square markers show the result of simulation obtained with flipped reconstruction events. Lower panels: on the left are the ratios of flipped ($\Ntptrt^{\mathrm{fl}}$) to direct ($\Ntptrt^{\mathrm{ev}}$) distribution in the data (markers) and in the simulation (dashed line); on the right is the ratio of the number of fake tracklets to the number of flipped tracklets.} \label{fig:fakes_eta}
\end{center}
\end{figure*}

The top left panel of Fig.~\ref{fig:fakes_eta} shows the pseudorapidity
distribution of tracklets reconstructed with \Mt\ and satisfying the
criteria of Eqs.~(\ref{eq:trackletcuts}) and~(\ref{eq:trackletcutmom}) 
in the 0--10\% centrality interval for data
(markers) and for the simulation (lines). 
The results of flipped reconstruction are also shown in the plot. 
The direct and flipped distributions are each similar between data and MC 
simulation but not identical, reflecting the fact that {\sc Hijing} does not reproduce 
the data in detail. However, the lower panel of Fig.~\ref{fig:fakes_eta} shows that 
the ratios of the distribution of the number of tracklets in the flipped and 
direct events are very similar between the data and the MC simulation. 
A breakdown of the MC simulation distribution into primary, secondary and fake tracklets contributions is
shown in the top right panel of Fig.~\ref{fig:fakes_eta}. The distribution
of $\Ntptrt^{\mathrm{fl}}(\eta)$, plotted with open markers, closely follows
the histogram of fake tracklets. The lower panel of the plot shows the ratio of fake tracklet distribution to the flipped distribution. This ratio is consistent with unity to within 5\% in the entire range of measured $\eta$. This agreement justifies the
subtraction of the fake tracklet contribution according to Eq.~(\ref{eq:flipsub})
for \Mt.  

Although the fake rate is largest in Method 2, the flipped method is used to estimate the rate directly from the data. In the 0--10\% centrality interval, the fake tracklet contribution estimated with this method amounts to 8\% of the yield at mid-pseudorapidity and up to 16\% at large pseudorapidity. In the same centrality interval, the fake tracklet contributions using \Mo and the pixel track method are smaller, vary from 2\% to 10\% and 0.2\% to 1.5\%, respectively, but are determined with MC. All three methods rely on the MC simulation to correct for the contribution of secondary particles.
\begin{figure*}[!htb]
\begin{center}
\includegraphics[width=0.49\textwidth]{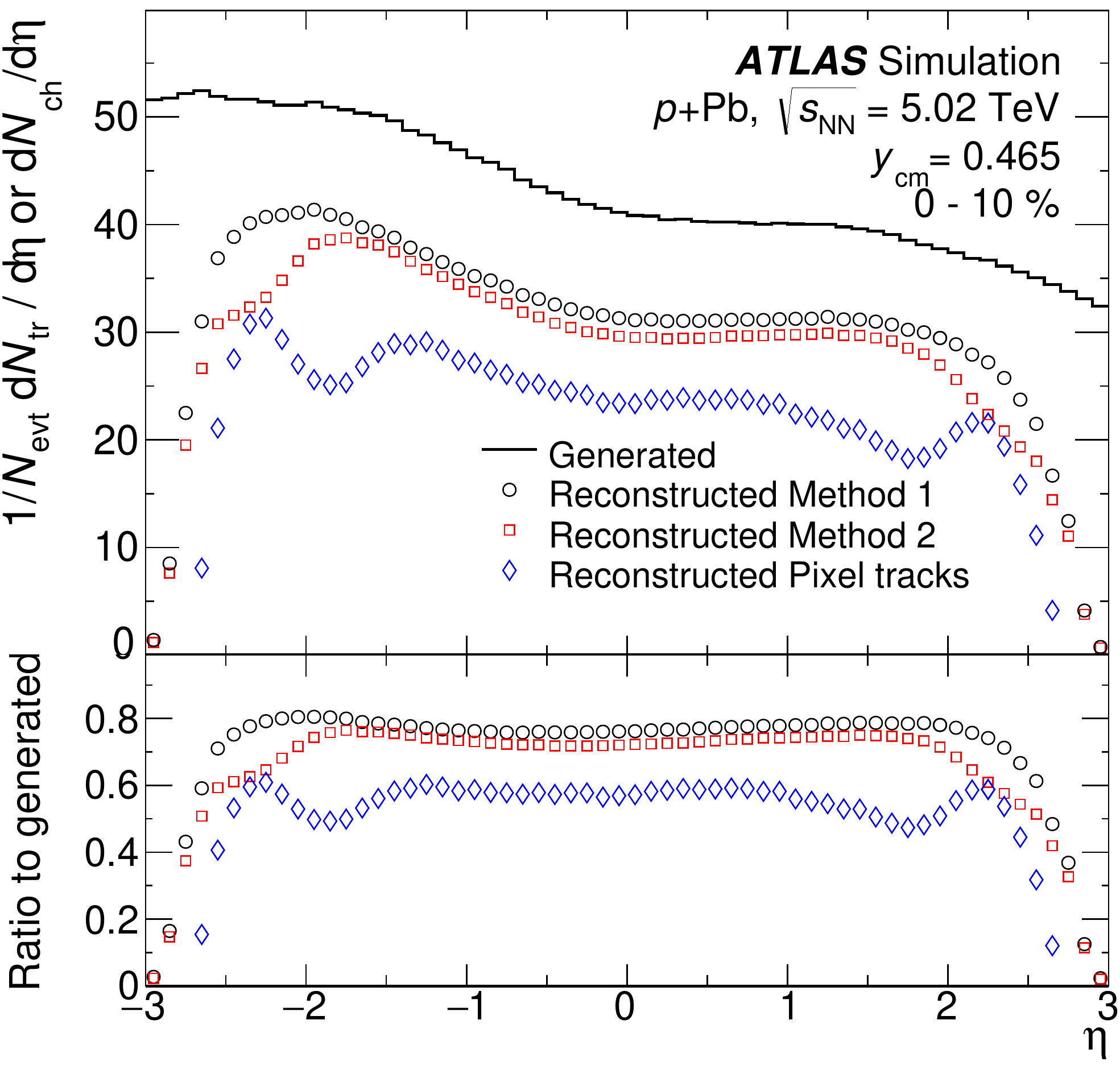} 
\includegraphics[width=0.49\textwidth]{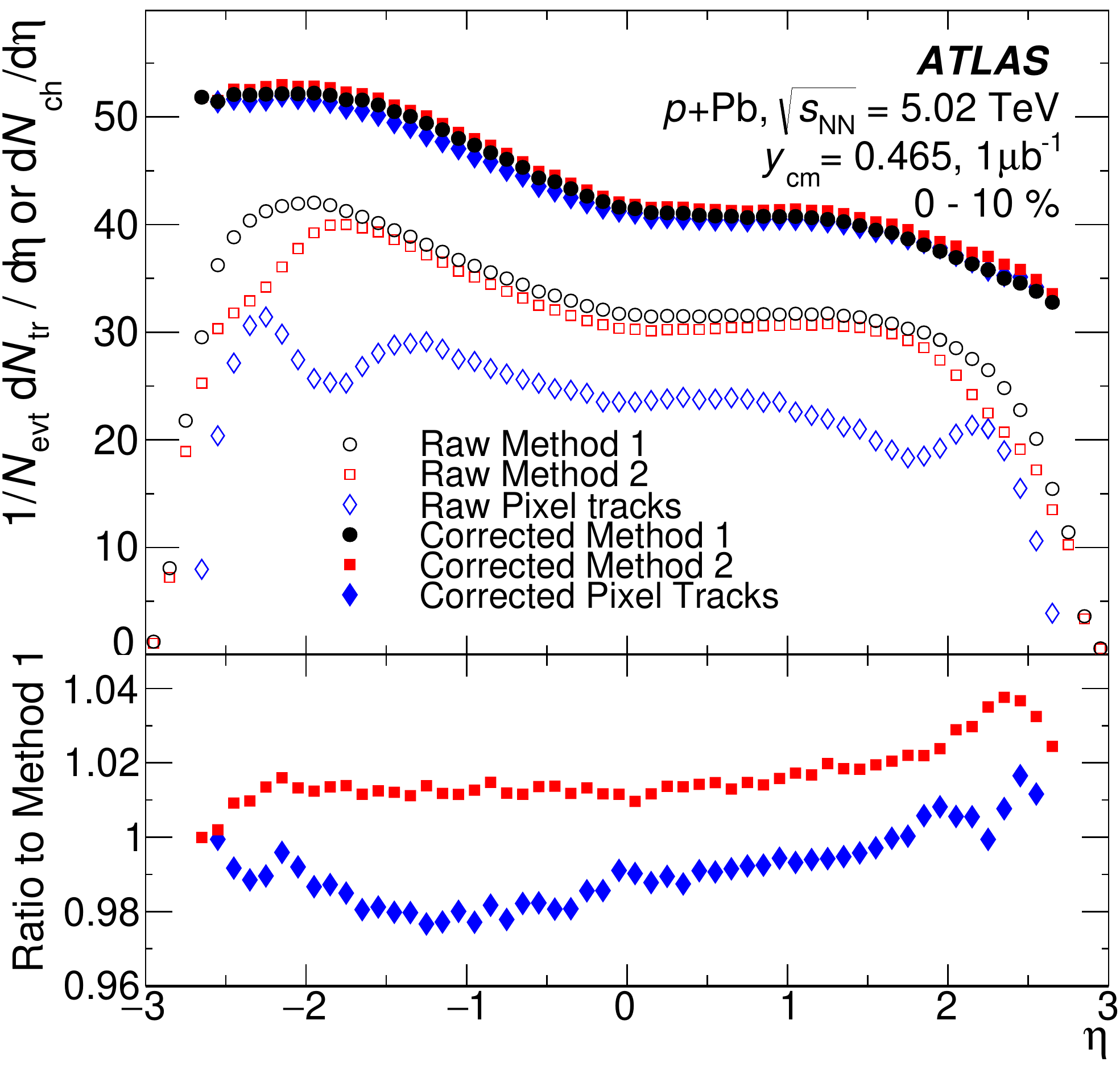}
\caption{Left top: distribution from the MC simulation for the generated number of primary charged particles per event ($\ensuremath{\dif N_{\text{ch}}}/\dif \eta$) shown with line, reconstructed number per event ($\ensuremath{1/N_{\text{evt}} \dif N_{\text{tr}}/\dif \eta}$) of tracklets from \Mo\ shown with circles, tracklets from \Mt\ after flipped event subtraction shown with squares, and pixel tracks shown with diamonds. Left bottom: the ratio of reconstructed to generated tracklets and pixel tracks. Right top: open markers represent the same $\ensuremath{1/N_{\text{evt}} \dif N_{\text{tr}}/\dif \eta}$ distributions as in the left panel, reconstructed in the data. Filled markers of the same shape represent corrected distributions corresponding to $\ensuremath{\dif N_{\text{ch}}/\dif \eta}$. Right bottom: The ratio of corrected distributions of \Mt\ and pixel tracks to \Mo.}
\label{fig:effic}
\end{center}
\end{figure*}

\subsection{Extraction of the charged-particle distribution}
The data analysis and corresponding corrections are performed in
eight intervals of detector occupancy (\occ) parameterised using the
number of reconstructed clusters in the first pixel layer and chosen to correspond to the eight \pPb\ centrality intervals, and in
seven intervals of \zvtx, each 50~mm wide. 
For each analysis method, a set of multiplicative correction factors is 
obtained from MC simulations according to 
\begin{equation}
C(\occ, \zvtx, \eta) \equiv \frac{N_{\mathrm{pr}}(\occ,\zvtx,\eta)}{N_{\mathrm{rec}}(\occ, \zvtx, \eta)}.
\label{eq:twopttrkcorr}
\end{equation}
Here, $N_{\mathrm{pr}}$ and $N_{\mathrm{rec}}$ represent the number of
primary charged particles at the generator level and the number of tracks or tracklets
at the reconstruction level, respectively. These correction factors
account for several effects: inactive areas in the
detector and reconstruction efficiency, contributions of residual
fake and secondary particles, and losses due to track or tracklet
selection cuts including particles with \pT\ below 0.1~\GeV. They
are evaluated as a function of \occ, \zvtx, and $\eta$ both for the
fiducial region, $\pT > 0.1$~\GeV, and for full 
acceptance, $\pT > 0$~\GeV. The results are presented in $\eta$-intervals 
of 0.1 unit width. Due to the excellent $\eta$-resolution of the tracklets, as seen from 
Fig.~\ref{fig:deta_phi_mc_data}, migration of tracklets between 
neighbouring bins is negligible.

The fully corrected, per-event charged-particle
pseudorapidity distributions are calculated according to
\begin{equation}
\frac{\mathrm{d}\Nch}{\mathrm{d}\eta} = 
\frac{1}{\Delta \eta} \frac{\sum \Delta \Nraw (\occ, \zvtx, \eta) C(\occ, \zvtx, \eta)}{\sum N_{\text{evt}}(\zvtx)},
\label{eq:aver}
\end{equation}
where $\Delta \Nraw $ indicates either the number of reconstructed
pixel tracks or two-point tracklets, 
$N_{\text{evt}}(\zvtx)$ is the number
of analysed events in the intervals of the primary vertex along the $z$ direction, and the 
sum in Eq.~(\ref{eq:aver}) runs over primary vertex intervals. 
The number of 
primary vertex intervals  
varies from seven for $|\eta|<2.2$ for two-point tracklets and $|\eta|<2$ for pixel
tracks to two at the edges of the measured pseudorapidity range of
$|\eta|<2.7$ for two-point tracklets and $|\eta|<2.5$ for pixel tracks respectively.
The primary vertex intervals used in the analysis are chosen such that 
$C(\occ, \zvtx, \eta)$ changes by less than 20\% between any pair of adjacent \zvtx intervals.

Figure~\ref{fig:effic} shows the effect of the applied correction for all three methods.
The left panels shows the MC simulation results based on {\sc Hijing}. The distribution
of generated primary charged particles is shown by a solid line and the
distributions of reconstructed tracks and tracklets are indicated by
markers in the upper left panel. The lower left panel shows the ratio
of reconstructed distributions to the generated distribution. Among the three
methods, the corrections for \Mo\ are the smallest, while the
pixel track method requires the largest corrections. 
The structure of the measured 
distribution for the pixel track
method around $\eta=\pm2$ is related to the transition between the
barrel and endcap regions of the detector. The open markers in the
right panel of Fig.~\ref{fig:effic} show the reconstructed
distribution from the data and the filled markers are the
corresponding distribution for the three methods after 
applying corrections derived from the simulation. 
The lower panel shows the ratio of the results obtained from \Mt\ and
the pixel track method to that obtained using \Mo. The three methods
agree within 2\% in the barrel region of the detector and
within 3\% in the endcap region. This agreement demonstrates that the rejection
of fake track or tracklets and the correction procedure are well understood. For this
paper, \Mo\ is chosen as the default result for \dndeta, \Mt\ is used
when evaluating systematic uncertainties, and the pixel track method is used
primarily as a consistency test, as discussed in detail below. 

\section{Systematic uncertainties}
\label{sec:syst}
The systematic uncertainties on the \dndeta\ measurement arise from three
main sources: inaccuracies in the simulated detector geometry,
sensitivity to selection criteria used in the analysis including the
residual contributions of fake tracklets and secondary particles, and differences
between the generated particles used in the simulation and the data. 
To determine the systematic uncertainties, the analysis is 
repeated in full for different variations of parameters or methods
and the results are compared to the standard \Mo\ results. A summary
of the results are presented in Table ~\ref{tbl:systerr}.

The uncertainty due to the simulated detector geometry arises 
primarily from the details of the pixel detector acceptance and efficiency. The
locations of the inactive pixel modules are matched between the data
and simulation. Areas smaller than a single module that are found to
have intermittent inefficiencies are estimated to contribute less
than 1.7\% uncertainty to the final result.  This uncertainty has no
centrality dependence, and is approximately independent of
pseudorapidity.  

The amount of inactive detector material in the tracking system is known with a precision 
of 5\% in the central region and up to 15\% in the forward region. 
In order to study the effect on the tracking efficiency, samples generated 
with increased material budget are used.  The net effect on the final result
is found to be 0.5--3\% independent of centrality. 

Uncertainties due to tracklet selection cuts are evaluated by
independently varying the cuts on $|\Delta\eta|$ and $|\Delta\phi|$ up
and down by 40\%. The effect of these variations is less than 1\%,
except at large values of $|\eta|$ where it is 1.5\%, and has only a weak
centrality dependence. 

 The systematic uncertainty due to applying the \pT\ reweighting procedure to the generated particles
 is taken from the difference in \dndeta\ between applying
 and not applying the reweighting procedure.  The uncertainty is
 less than 0.5\% 
 for $|\eta|<1.5$ and grows to 3.0\% towards the edges of the $\eta$
 acceptance. The uncertainty has a centrality dependence because
 the \pt\ distributions in central and peripheral collisions are
 different. 

Tracklets are reconstructed using \Mo\ for particles with
$\pT>0.1$~\GeV.  The unmeasured region of the spectrum contributes
approximately 6\% to the final \dndeta distribution. The systematic uncertainty on
the number of particles with $\pT\leq 0.1$~\GeV\ is partially included in the variation of
the tracklet $\Delta\phi$ selection criteria. An additional uncertainty
is evaluated by varying the shape of the
spectra below 0.1~\GeV. This uncertainty is estimated to be as much as
2.5\% at large values of $|\eta|$ and has a weak centrality dependence. 

To test the sensitivity to the particle composition in {\sc Hijing}, 
the fraction of pions, kaons and protons in {\sc Hijing} are varied within a 
range based on measured differences in particle composition
between \pp\ and \PbPb\ collisions \cite{Chatrchyan:2013eya,Abelev:2012wca}. 
The resulting changes in \dndeta\ are found to be less than 1\% for
all centrality intervals. 
 
Systematic uncertainties due to the fake tracklets are estimated
by comparing the results of the two tracklet methods. The
differences in the most central collisions are found to vary with pseudorapidity 
from 1.5\% in the barrel region to about 2.5\% at the ends
of the measured pseudorapidity range. 

\begin{table}[!htb]
\begin{center}
\begin{tabular}{l|cc|cc}
                                          & \multicolumn{2}{c|}{60--90\% centrality} & \multicolumn{2}{c}{0--1\%  centrality}\\
Source                                 & barrel & endcap & barrel & endcap\\
\hline
Inactive modules                    &\multicolumn{2}{c|}{1.7\%}          &\multicolumn{2}{c}{1.7\%} \\ 
Extra material                               & 0.5\%     & 3.0\%                         & 0.5\%     & 3.0\% \\ 
Tracklet selection                          & 0.5\%   & 1.5\%                           & 0.5\%   & 1.5\% \\ 
\pt\ reweighting                          & 0.5\% & 0.5\%                           & 0.5\%  & 3.0\% \\
Particles with $\pT\leq0.1$~\GeV\ & 1.0\%     & 2.5\%                 & 1.0\%   & 2.0\% \\ 
Particle composition                     &\multicolumn{2}{c|}{1.0\%}         & \multicolumn{2}{c}{1.0\%} \\ 
Contribution of fake tracklets        & 1.5\% & 2.0\%                             & 1.5\% & 2.5\%   \\
Event selection efficiency	             & 5.0\%    & 6.0\%                          & 0.5\%  & 0.5\% \\   
\hline
Total				            & 5.7\%    & 7.9\%                           & 2.9\%  & 5.9\% \\
\end{tabular}
\caption{Summary of the various sources of systematic uncertainty and their estimated impact on the \dndeta\ measurement in central (0--1\%) and peripheral (60--90\%) \pPb\ collisions.}
\label{tbl:systerr}
\end{center}
\end{table}
The uncertainty associated with the event selection efficiency for the fiducial class of \pPb\ events
is evaluated by defining new \sumETPb\ centrality ranges after
accounting for an increase (decrease) in the efficiency
by 2\% and repeating the full analysis. This resulting change of
the \dndeta distribution is less than 0.5\% in central collisions; it increases to
6\% in peripheral collisions. 

The uncertainties from each source were evaluated separately in each centrality and pseudorapidity to allow for their partial or complete cancellation in the ratios of \dndeta distributions. 
The impact in
 different regions of pseudorapidity and centrality are shown in
 different columns of Table~\ref{tbl:systerr}. Uncertainties coming from different sources 
 and listed in the same column are treated as
 independent. The resulting total systematic uncertainty shown 
 in the lower line of the table is the sum in quadrature of the individual contributions.

%% file: 6_results.tex
\begin{figure*}[!htb]
\begin{center}
\includegraphics[width=0.49\textwidth, angle=0]{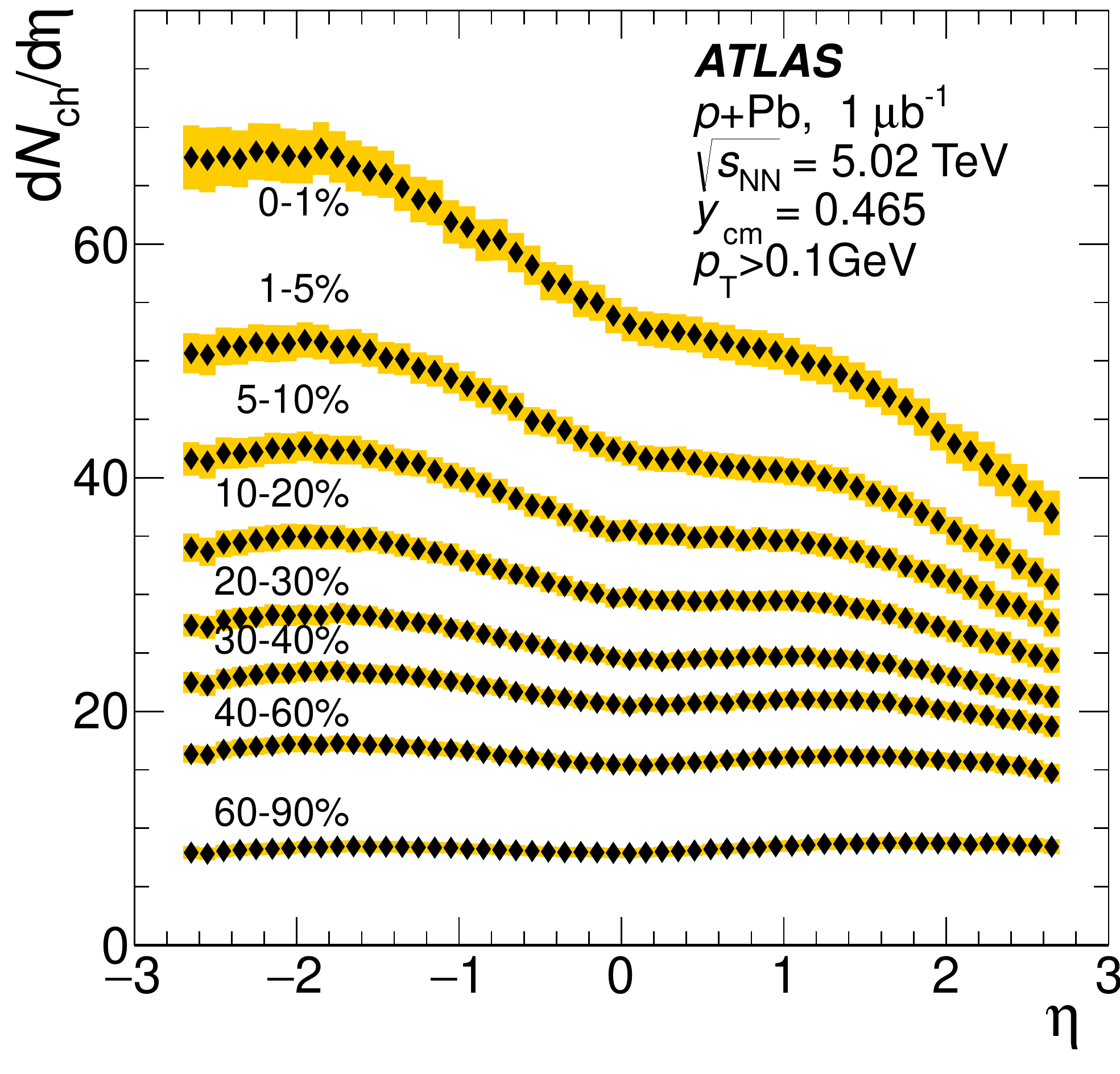}
\includegraphics[width=0.49\textwidth, angle=0]{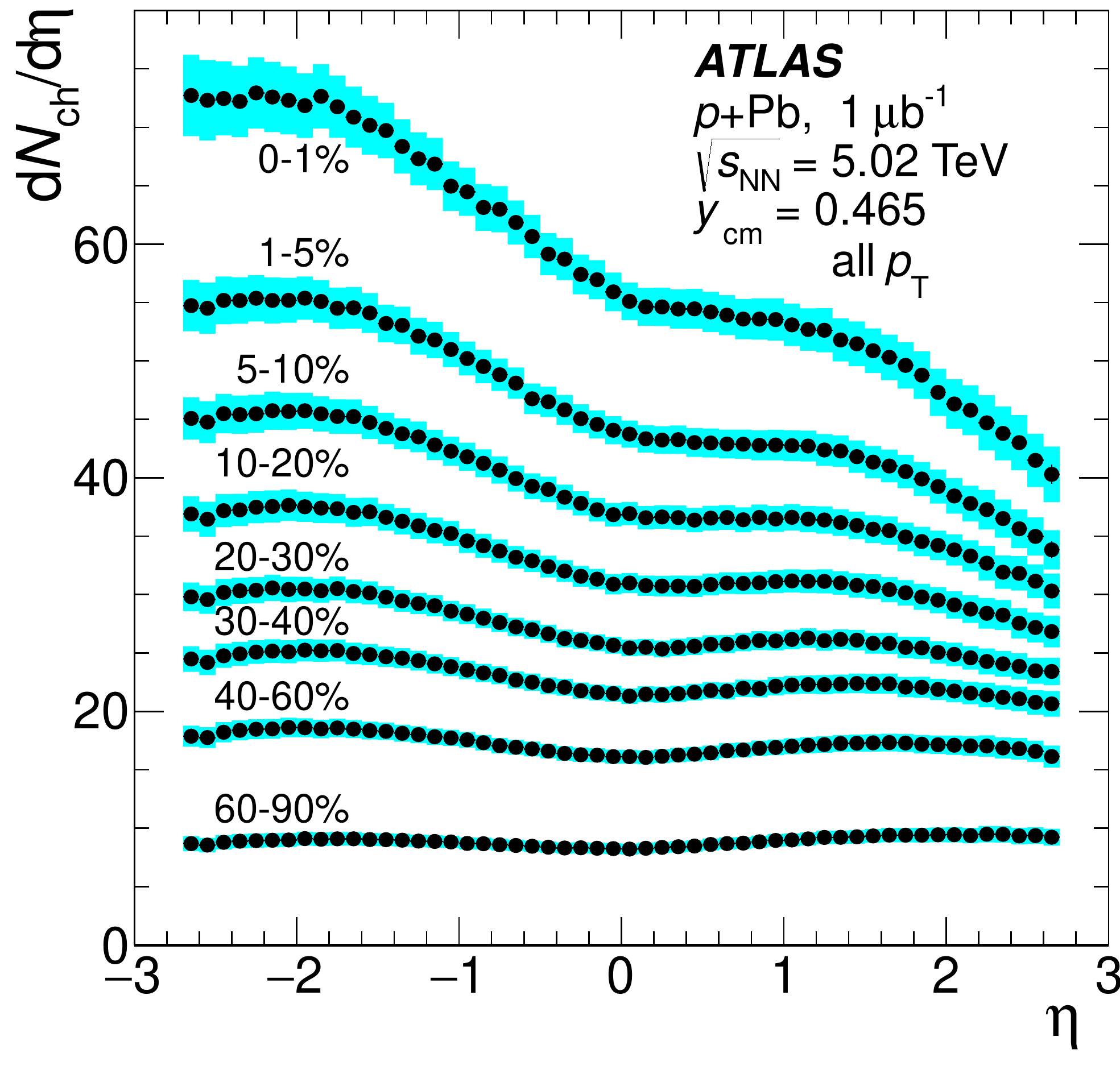}
\caption{Charged-particle pseudorapidity distribution measured in several centrality intervals. Left: \dndeta for charged particles with $\pT>0.1$~\GeV. Right: \dndeta for charged particles with $\pT>0$~\GeV. Statistical uncertainties, shown with vertical bars, are typically smaller than the marker size. Shaded bands indicate systematic uncertainties on the measurements. }
\label{fig:combine}
\end{center}
\end{figure*}

Figure~\ref{fig:combine} presents the charged-particle 
pseudorapidity distribution \dndeta\ for \pPb\ collisions at 
\sqn=5.02\,\TeV\ in the pseudorapidity interval $|\eta|<2.7$ 
for several centrality intervals. 
The left panel shows the \dndeta distribution measured in the fiducial acceptance of the ATLAS detector, 
detecting particles with $\pT>0.1$~\GeV. The results for the \dndeta distribution with $\pT>0$~\GeV\ are shown in the right panel of Fig.~\ref{fig:combine}. 
The charged-particle pseudorapidity distribution increases by typically 5\%, consistent with 
extrapolation of spectra measured in \pp\ collisions to zero \pT~\cite{Aad:2010ac}. 
At the edges of the measured pseudorapidity interval, it increases \dndeta by 11\%.

\begin{figure}[!h]
\begin{center}
\includegraphics[width=0.49\textwidth, angle=0]{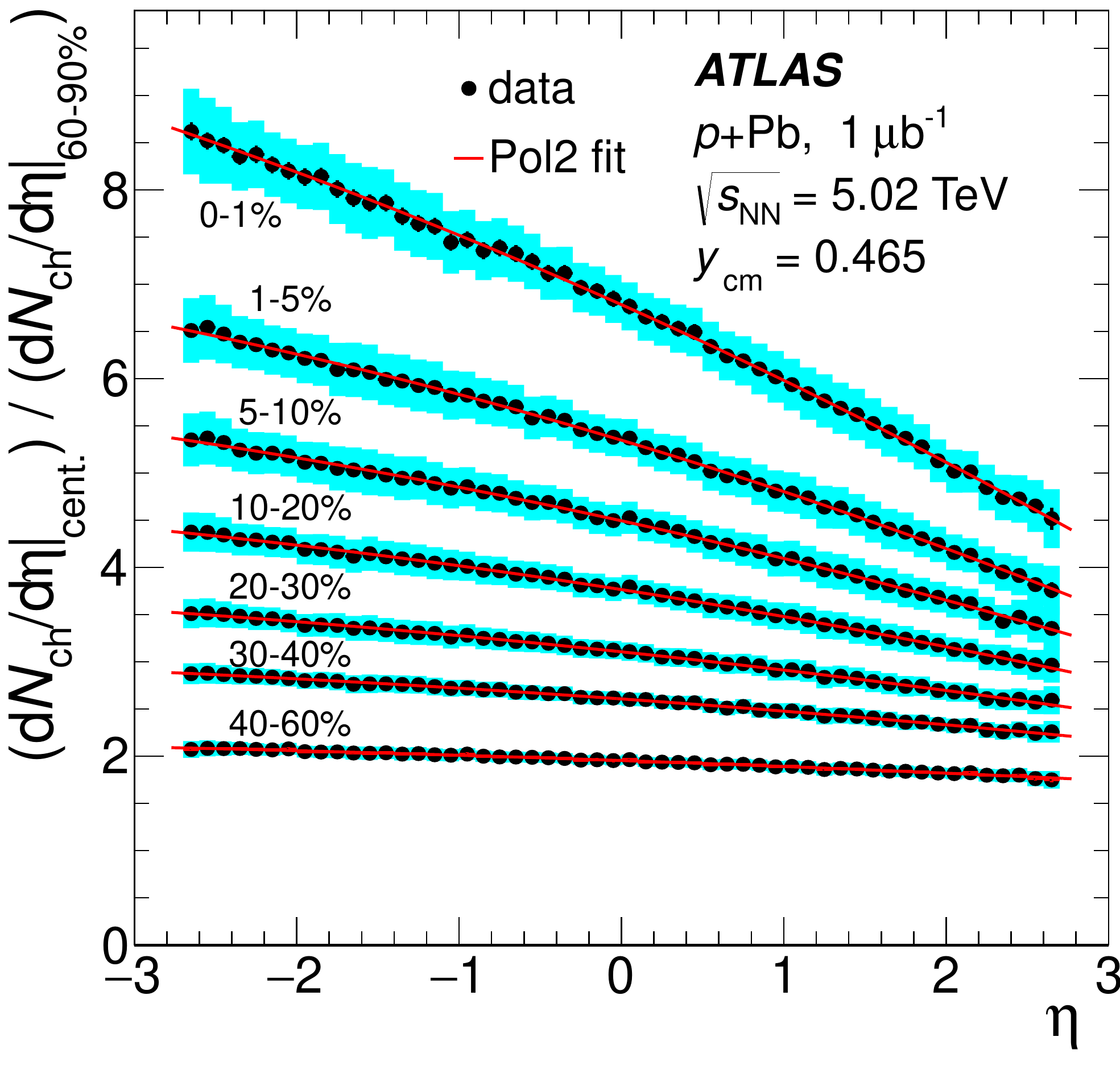}
\caption{Ratios of \dndeta distributions measured in different centrality intervals to that in the peripheral (60--90\%) centrality interval. Lines show the results of second-order polynomial fits to the data points.} 
\label{fig:ratio_to_peripheral}
\end{center}
\end{figure}

In the most peripheral collisions with a centrality of 60--90\%, the \dndeta
distribution has a doubly-peaked shape similar to that seen in
\pp\ collisions~\cite{Aad:2010ac,Aad:2010rd}.
In collisions that are more central, 
the shape of \dndeta becomes progressively more asymmetric, with more
particles produced in the Pb-going direction than in the proton-going
direction. To investigate further the centrality evolution, the 
\dndeta distributions in each centrality interval are divided by the 
\dndeta distribution for the 60--90\% interval. 
The results are shown in Fig.~\ref{fig:ratio_to_peripheral}, where 
the double-peak structure disappears in the ratios. The ratios are
observed to grow nearly linearly with decreasing pseudorapidity, with a slope
whose magnitude increases from peripheral to central collisions. 
In the 0--1\% centrality interval, the ratio changes by almost a
factor of two over the measured $\eta$-range. The greatest increase in
multiplicity between adjacent centrality intervals occurs between the 1--5\%
and 0--1\% intervals. Averaged over the $\eta$-interval of the
measurement, the \dndeta distribution increases by more than 25\% between
the 1--5\% and 0--1\% intervals. 

\begin{figure}[!htb] 
\begin{center} 
\includegraphics[width=0.49\textwidth, angle=0]{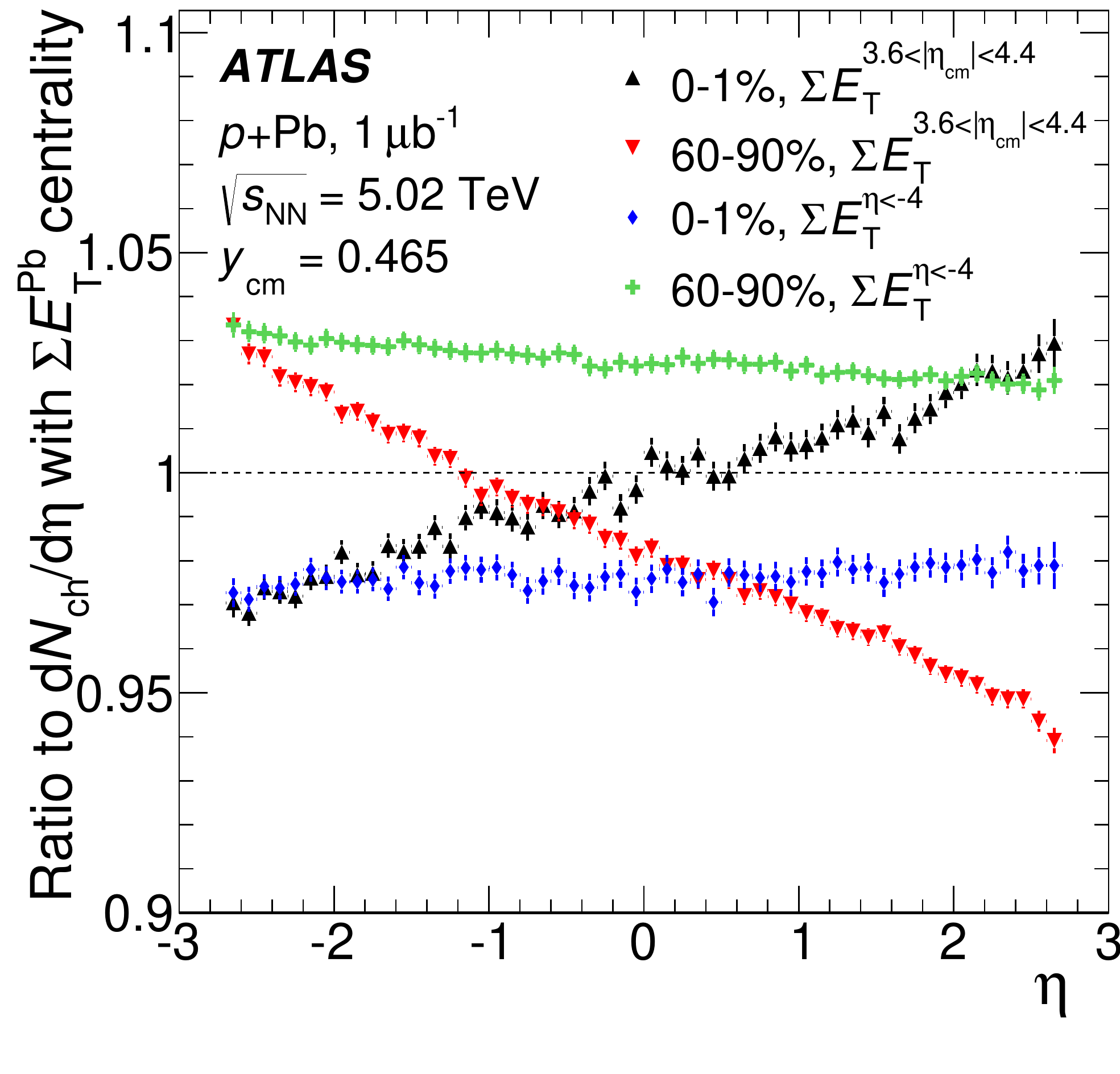} 
\caption{Ratios of \dndeta obtained using alternative centrality definitions to the nominal results presented in this paper as a function of $\eta$ for the 0--1\% and 60--90\% centrality bins.} 
\label{fig:altcentratios} 
\end{center} 
\end{figure} 

In addition to the results presented in
Figs.~\ref{fig:combine} and~\ref{fig:ratio_to_peripheral},
the \dndeta measurement is repeated using the alternative
definitions of the event centrality variables defined in
Sect.~\ref{sec:centrality_def}. Figure~\ref{fig:altcentratios}
demonstrates the sensitivity of the measured \dndeta to the choice of
centrality variable by showing the ratios of the \dndeta distributions in
the most central and most peripheral intervals under the \sumETPbfar\
and \sumETsymm\ centrality definitions to those obtained with the
nominal \sumETPb\ definition. Using the
\sumETPbfar\ centrality definition, the \dndeta distributions change
in an approximately $\eta$-independent fashion by $-3$\% and $+3$\%
for the 0--1\% and 60--90\% intervals, respectively. The \dndeta
distributions in the other centrality intervals change in a manner
that effectively interpolates between these extremes. As a result,
the increase in \dndeta between the most peripheral and
most central collisions would be reduced by 6\% 
relative to the nominal
measurement. Using the symmetric, \sumETsymm\ centrality definition,
the \dndeta distribution in each interval 
changes in an $\eta$-dependent way such that the ratio is consistent
with a linear function of $\eta$. 
The change is at most 6\% at the ends of the 
$\eta$ range in the most central and most peripheral centrality
intervals, and smaller elsewhere. Thus, for the symmetric centrality
selection the ratios in Fig.~\ref{fig:ratio_to_peripheral} for the
0--1\% bin would increase by 9\% at $\eta =
2.7$, and decrease by 6\% at $\eta = -2.7$. Generally, the alternative centrality
definitions considered in this analysis yield no qualitative 
and only modest quantitative changes in the centrality dependence of
the \dndeta\ distributions. These variations should not be considered a
systematic uncertainty on the \dndeta\ measurement but do indicate that the
particular centrality method used in the analysis must be accounted
for when interpreting the results of the measurement.

\begin{figure}[!htb]
\begin{center}
\includegraphics[width=0.49\textwidth]{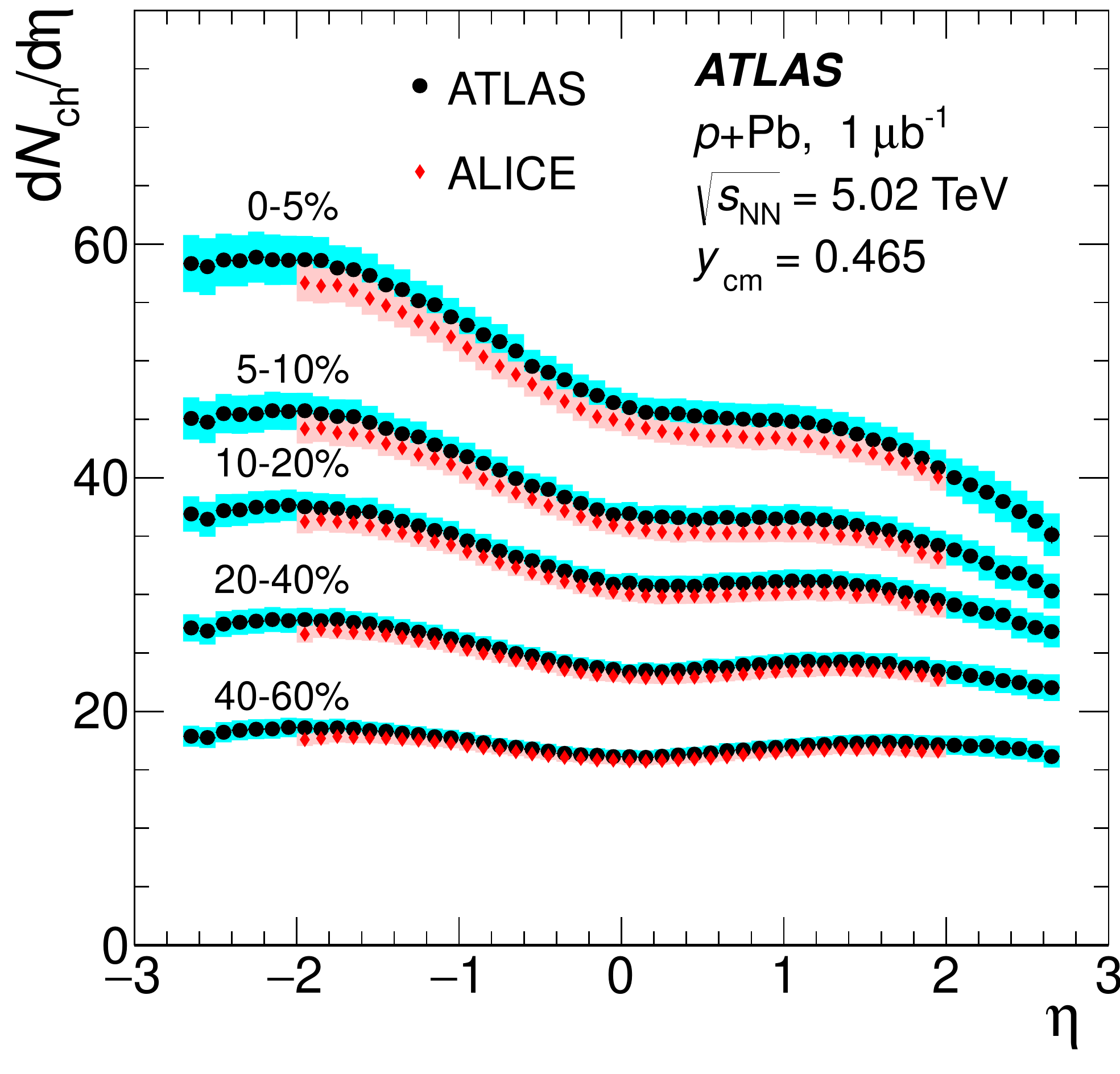}
\caption{Charged-particle pseudorapidity distribution \dndeta measured in different centrality intervals compared to similar results from the ALICE experiment~\cite{Adam:2014qja} using the ``{\tt V0A}'' centrality selection. The ATLAS centrality intervals have been combined, where possible, to match the ALICE centrality selections.}
\label{fig:alice}
\end{center}
\end{figure}

Figure~\ref{fig:alice} shows a comparison, where possible, of the measurements
presented in this paper to results from the ALICE experiment~\cite{Adam:2014qja} 
using a centrality definition that
is based on the detector covering the pseudorapidity region $-5.1<\eta<-2.8$, 
similar to the \sumETPb-based selection used in this
measurement. The ATLAS results for 0--1\% and 1--5\% centrality intervals are 
combined to match the ALICE experiment result for 0--5\% interval. Similarly, the
20--30\% and 30--40\% intervals are combined to match the ALICE experiment result for
20--40\% interval. The results from the two experiments are consistent
with each other.

\section{Particle multiplicities per participant pair}
\label{sec:perNpartyields}
\begin{figure}[!htb]
\begin{center}
\includegraphics[width=0.49\textwidth, angle=0]{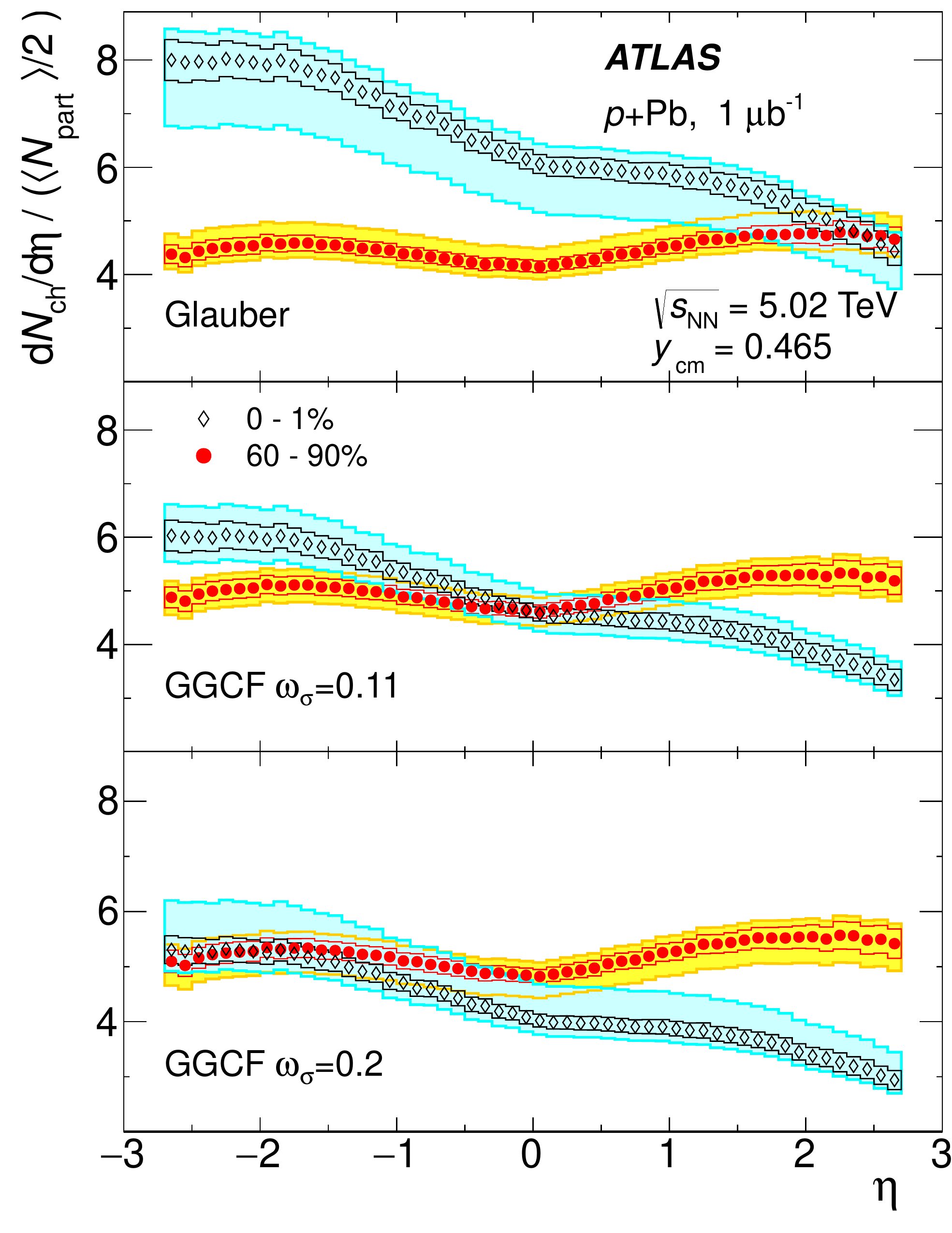}
\caption{Charged-particle pseudorapidity distribution \dndeta  per pair of participants as a function of \eta\ for 0--1\% and 60--90\% centrality intervals for the three models used to calculate \Npart. The standard Glauber calculation is shown in the top panel, the GGCF model with $\omega_{\sigma}=0.11$ in the middle and $\omega_{\sigma}=0.2$ 
in the lowest panel. The bands shown with thin lines represent the systematic uncertainty of the \dndeta measurement, the shaded bands indicate the total systematic uncertainty including the uncertainty on  \avgNpart. Statistical uncertainties, shown with vertical bars, are typically smaller than the marker size.}
\label{fig:eta_per_part}
\end{center}
\end{figure}

A common way of representing the centrality dependence of particle
yields in \NucNuc\ and \pA\ collisions is by showing the yield per
participant or per participant pair, $\avgNpart/2$, which is determined 
for each centrality interval and each geometrical model as shown in Fig.~\ref{fig:fitnpart}. 
Figure~\ref{fig:eta_per_part} shows \dndeta per participant pair for
the most central and most peripheral intervals of centrality measured 
as a function of \eta\ for three different
models of the collisions geometry: the standard Glauber model and
the GGCF model with $\omega_{\sigma} = 0.11$ and 0.2 in the top,
middle and lower panels, respectively.  
The results for the most peripheral (60--90\%) centrality interval,
shown with circles, are similar between all
three panels. This is due to relatively small difference between the
calculations of \avgNpart\ 
for Glauber and GGCF models in this centrality interval. The shape of the  
distribution indicates more abundant particle production in the
proton-going direction in comparison to the Pb-going. 
This can be explained by the higher energy of the proton compared to the energy 
of a single nucleon in the lead nucleus in the laboratory system. 
In the most central
collisions (0--1\%), shown with diamond markers in all three panels,
this trend  is reversed. Conversely, the magnitude of \dndeta per participant pair strongly
depends on the geometric model used to calculate \avgNpart. The
point at which the central and peripheral scaled distributions cross
each other also depends on the choice of geometric model.  

\begin{figure}[!htb] 
\begin{center}
\includegraphics[width=0.49\textwidth, angle=0]{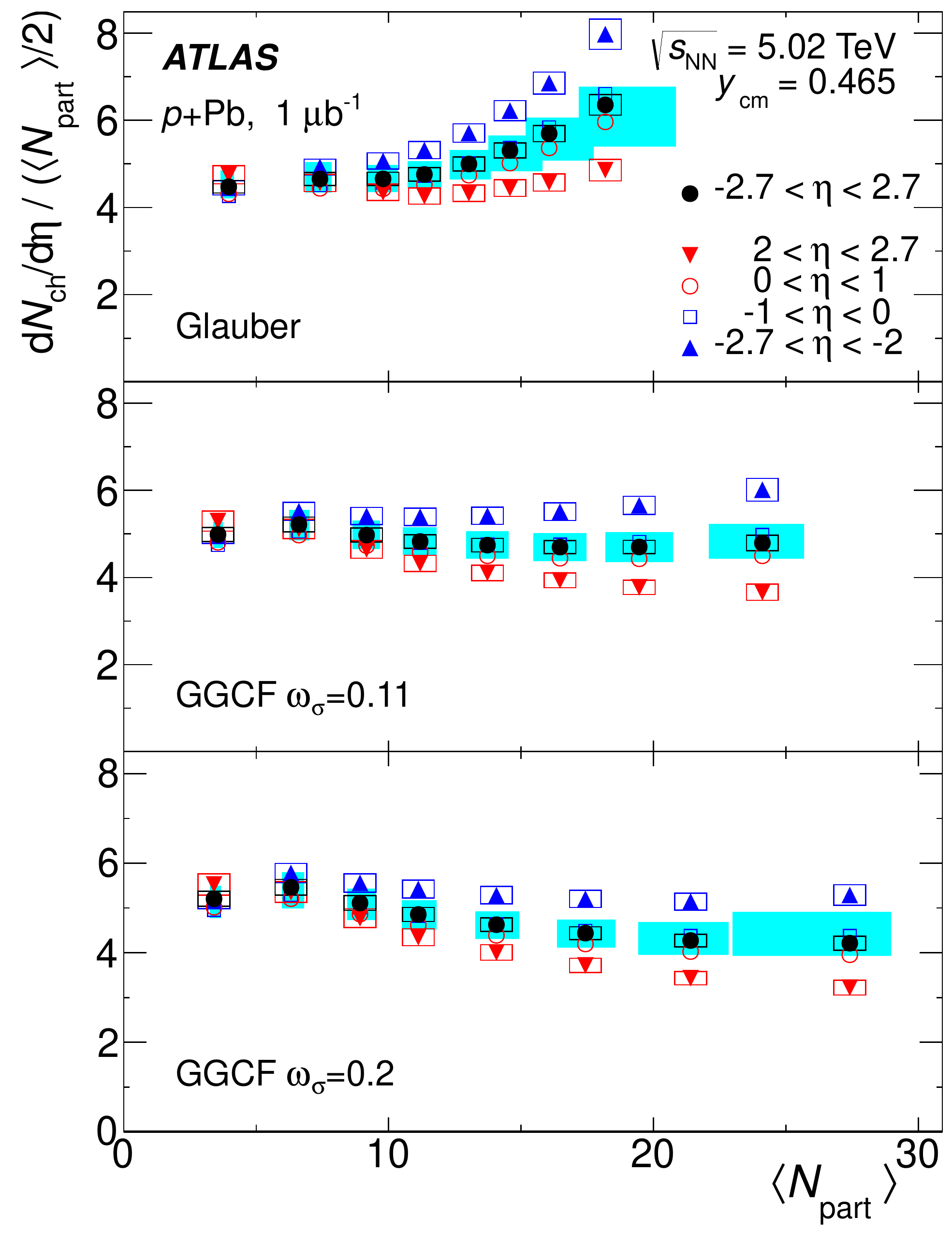}
\caption{Charged-particle pseudorapidity distribution \dndeta per pair of participants as a function of \avgNpart\ in several $\eta$-regions for the three models of the geometry: the standard Glauber model (top panel), the GGCF model with  $\omega_{\sigma}=0.11$ (middle panel) and GGCF with $\omega_{\sigma}=0.2$ (bottom panel). The open boxes represent the systematic uncertainty of the \dndeta measurement only, and the width of the box is chosen for better visibility (they are not shown for $-1.0<\eta<0$ and $0<\eta<1$). The shaded boxes represent the total uncertainty (they are shown only on $-2.7<\eta<2.7$ interval for visibility) which is dominated by the uncertainty of the \avgNpart\ given in Table~\ref{tbl:npartvalues} and Fig.~\ref{fig:fitnpart}. This uncertainty is asymmetric due to the asymmetric uncertainties on \avgNpart. The statistical uncertainties are smaller than the marker size for all points.} 
\label{fig:npart}
\end{center}
\end{figure}

Figure~\ref{fig:npart} shows the \dndeta distribution per participant pair 
as a function of \avgNpart\ for 
the three different models of the collisions geometry. Since the charged-particle 
yields have significant pseudorapidity dependence, 
$\dndeta/(\avgNpart/2)$ is presented in five $\eta$ intervals
including the full pseudorapidity interval, $-2.7<\eta<2.7$. In the
region $0<\eta<1$, the \dndeta distribution is consistent with an empirical fit to 
inelastic \pp\ data that suggest \dndeta increases with
centre-of-mass energy, $\sqrt{s}$, as $\left(\propto
s^{0.10}\right)$~\cite{ALICE:2012xs}. 

The \dndeta/(\avgNpart/2) values from the standard Glauber model are
approximately constant up to $\avgNpart \approx 10$ and then increase
for larger \avgNpart. This trend is absent in the GGCF model
with $\omega_{\sigma}=0.11$, which shows a relatively constant
behaviour for the integrated yield divided by the number of
participant pairs.  The $\dndeta/(\avgNpart/2)$ values from the
GGCF model with $\omega_{\sigma}=0.2$ show a slight decrease
with \avgNpart\ in all $\eta$ intervals. 

The presence or absence of \avgNpart\ scaling does not 
suggest a preference for one or another of the geometric
models. However, this study emphasises that considering 
fluctuations of the nucleon--nucleon cross-section in the
GGCF model may lead to significant changes in the \Npart\
scaling behaviour of the \pPb\ \dndeta data and, thus, their
interpretations.

%% file: 7_conclusions.tex
This \paper\ presents a measurement of the centrality dependence
of the charged-particle pseudorapidity distribution, \dndeta, measured in 
approximately 1 $\mu$b$^{-1}$ of \pPb\ collisions at a nucleon--nucleon centre-of-mass energy
of \sqn=5.02\,\TeV\ collected by the ATLAS detector at the LHC.
The fully corrected measurements are presented for the fiducial
acceptance of the ATLAS detector ($\pT > 0.1$~\GeV) and in the full acceptance ($\pT >0$~\GeV). 
The \dndeta distributions are
presented as a function of pseudorapidity over the range
$-2.7<\eta<2.7$ and as a function of collision centrality for the 0--90\%
\pPb\ collisions. The centrality is characterised using the energy deposited in 
the forward calorimeter covering $-4.9 < \eta < -\FCalLowEta$ in the
Pb-going direction.  

The shape of \dndeta evolves gradually with centrality from an
approximately symmetric shape in the most peripheral collisions to a
highly asymmetric distribution in the most central collisions. 
The ratios of \dndeta measured in different centrality intervals 
to the \dndeta distribution in the most peripheral interval are approximately
linear in $\eta$ with a slope that is strongly dependent on
centrality. It is noteworthy that the greatest increase in
charged-particle multiplicity between successive centrality bins
occurs between the 1--5\% and 0--1\% centrality bins.

The results are also interpreted using models of the underlying
collision geometry. The average number of participants in each
centrality interval, \avgNpart, is estimated using a standard
Glauber model Monte Carlo simulation with a fixed nucleon--nucleon
cross-section, as well as with two Glauber--Gribov colour fluctuation
models which allow the nucleon--nucleon cross-section to fluctuate
event-by-event.  The \Npart\ dependence of $\dndeta/(\avgNpart/2)$ is found to be
sensitive to the modelling of the \pPb\ collision geometry, especially
in the most central collisions: while the standard Glauber modelling
leads to a strong increase in the multiplicity per participant pair
for collisions in the centrality range (0--30)\% the GGCF
model produces a much milder centrality dependence.  

These results point to the importance of understanding not just the
initial state of the nuclear wave function, but also the fluctuating
nature of nucleon--nucleon collisions themselves.

%% file: Acknowledgements2015-09-15.tex

We honour the memory of our colleague Alexey Antonov, who was closely involved in the work described here, and died shortly after its completion.

We thank CERN for the very successful operation of the LHC, as well as the
support staff from our institutions without whom ATLAS could not be
operated efficiently.

We acknowledge the support of ANPCyT, Argentina; YerPhI, Armenia; ARC, Australia; BMWFW and FWF, Austria; ANAS, Azerbaijan; SSTC, Belarus; CNPq and FAPESP, Brazil; NSERC, NRC and CFI, Canada; CERN; CONICYT, Chile; CAS, MOST and NSFC, China; COLCIENCIAS, Colombia; MSMT CR, MPO CR and VSC CR, Czech Republic; DNRF, DNSRC and Lundbeck Foundation, Denmark; IN2P3-CNRS, CEA-DSM/IRFU, France; GNSF, Georgia; BMBF, HGF, and MPG, Germany; GSRT, Greece; RGC, Hong Kong SAR, China; ISF, I-CORE and Benoziyo Center, Israel; INFN, Italy; MEXT and JSPS, Japan; CNRST, Morocco; FOM and NWO, Netherlands; RCN, Norway; MNiSW and NCN, Poland; FCT, Portugal; MNE/IFA, Romania; MES of Russia and NRC KI, Russian Federation; JINR; MESTD, Serbia; MSSR, Slovakia; ARRS and MIZ\v{S}, Slovenia; DST/NRF, South Africa; MINECO, Spain; SRC and Wallenberg Foundation, Sweden; SERI, SNSF and Cantons of Bern and Geneva, Switzerland; MOST, Taiwan; TAEK, Turkey; STFC, United Kingdom; DOE and NSF, United States of America. In addition, individual groups and members have received support from BCKDF, the Canada Council, CANARIE, CRC, Compute Canada, FQRNT, and the Ontario Innovation Trust, Canada; EPLANET, ERC, FP7, Horizon 2020 and Marie Skłodowska-Curie Actions, European Union; Investissements d'Avenir Labex and Idex, ANR, Region Auvergne and Fondation Partager le Savoir, France; DFG and AvH Foundation, Germany; Herakleitos, Thales and Aristeia programmes co-financed by EU-ESF and the Greek NSRF; BSF, GIF and Minerva, Israel; BRF, Norway; the Royal Society and Leverhulme Trust, United Kingdom.

The crucial computing support from all WLCG partners is acknowledged
gratefully, in particular from CERN and the ATLAS Tier-1 facilities at
TRIUMF (Canada), NDGF (Denmark, Norway, Sweden), CC-IN2P3 (France),
KIT/GridKA (Germany), INFN-CNAF (Italy), NL-T1 (Netherlands), PIC (Spain),
ASGC (Taiwan), RAL (UK) and BNL (USA) and in the Tier-2 facilities
worldwide.

%% file: 5_glauber.tex
\begin{figure}[hb!]
\centerline{
\includegraphics[width=0.60\textwidth]{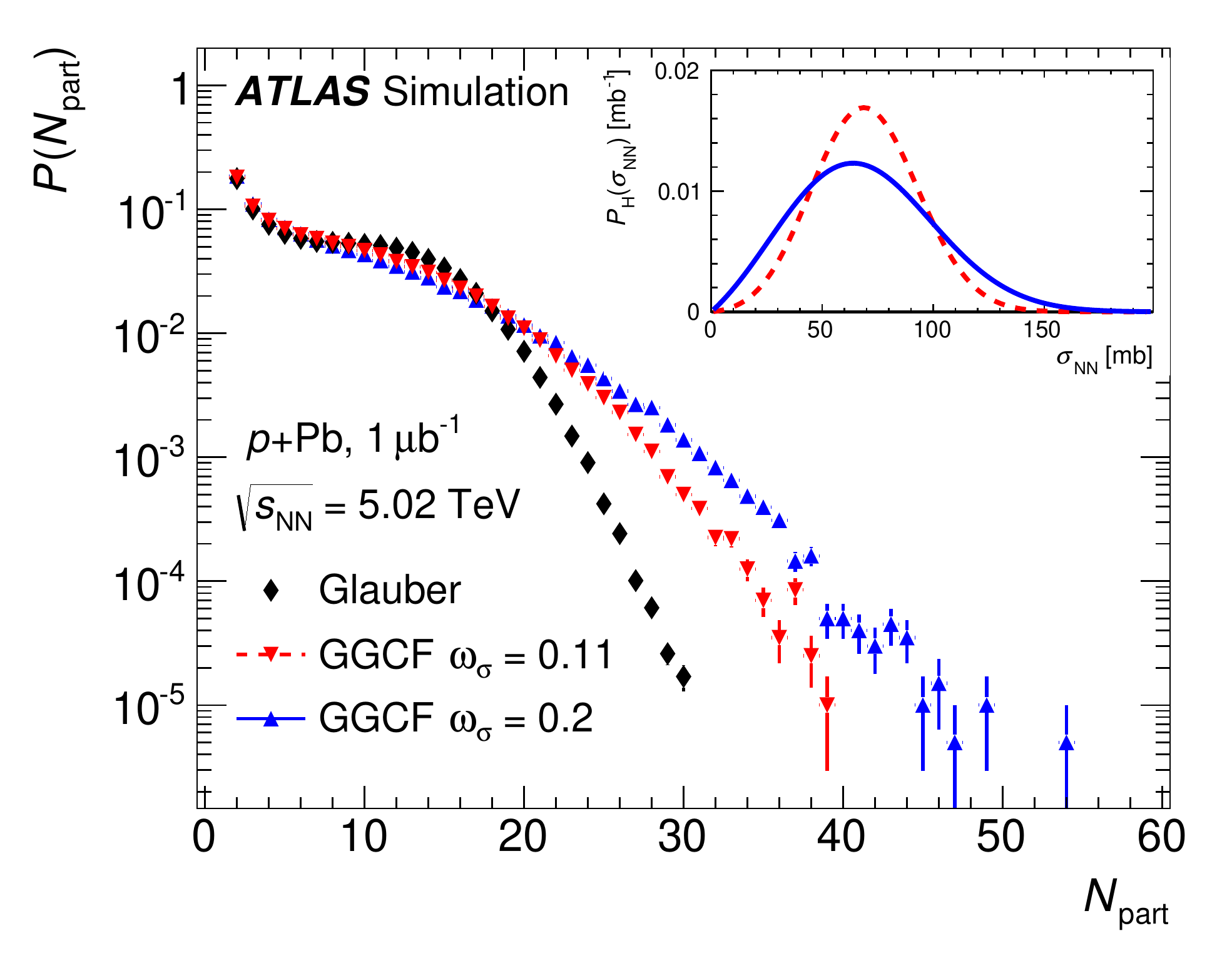}
}
\caption{Glauber and GGCF \Npart\ distributions for 5.02~\TeV\ \pPb\ collisions obtained from one million simulated events each. The inset shows the GGCF $P_{\text{H}}(\sigNN)$\ distributions for $\omsig= 0.11$ and 0.2.}
\label{fig:GG}
\end{figure}

\hyphenation{Glau-ber deu-te-ron}
The PHOBOS Glauber MC program~\cite{Alver:2008aq} is used to perform
the standard Glauber model calculations used in this analysis. The
Pb nucleon density is taken to be a Woods--Saxon distribution with
radius and skin depth parameters, $R = 6.62$~fm and $a =
0.546$~fm~\cite{DeJager:1987qc}, respectively.  The nucleon--nucleon 
inelastic cross-section is taken to be 70~mb. The resulting probability distribution, 
$P(\Npart)$, of the number of participating nucleons \Npart\ 
-- nucleons that undergo at least one
hadronic scattering during the \pPb\ collision -- is shown in
Fig.~\ref{fig:GG}.  

The GGCF model is implemented in a modified version of the PHOBOS MC program.
Following Ref.~\cite{Guzey:2005tk}, the probability distribution 
to find the nucleons in a configuration having a nucleon--nucleon
scattering cross-section, $\sigma$, is taken to be  
\begin{equation}
P ( \sigma) = \rho \left(\frac{ \sigma }{ \sigma + \sigz } \right) \exp \left\{ - \frac{ (\sigma / \sigz - 1)^2 }{ \Omega^2 } \right\}.
\label{eq:Psigtot}
\end{equation}
Here, $\rho$ is a normalisation constant, \OCap\ controls the width
of the $P(\sigma)$\ distribution, and \sigz\ determines the
configuration-averaged total cross-section
$\sigtot \equiv \langle \sigma \rangle$. 
The inelastic cross-section, \sigNN, is taken to be
a constant fraction, $\lambda$, of the total cross-section \cite{Alvioli:2013vk}
so the probability distribution of \sigNN\ is given by 
\begin{equation}
P_{\text{H}}(\sigNN) = \frac{1}{\lambda} P(\sigNN/\lambda).
\label{eq:PsigNN}
\end{equation}

The values used in this analysis for $\Omega$, \sigz, \sigtot\ and
$\lambda$ corresponding to $\omsig = 0.11, 0.2$ are shown in
Table~\ref{tbl:CFParam}. 
The first, earlier analysis yielding $\omsig =
0.11$ \cite{Guzey:2005tk} assumed
$\sigtot = 86$~mb, consistent with the Donnachie and Landshoff \cite{Donnachie:1992ny}
parameterisation of $\sigtot\left(s\right)$. The 
second analysis yielding $\omsig = 0.2$ used an updated 
measurement of the \pp\ total cross-section at the LHC \cite{PhysRevLett.111.012001} to set $\sigtot =
94.8$~mb. However, modifying the parameters for the $\omsig = 0.11$ case to be
consistent with this improved knowledge of \sigtot\ produces a negligible
change in the resulting $P(\sigma)$ distribution.
\begin{table}[b]
\centerline{
\begin{tabular}{l|r|r}
Parameter & $\omsig = 0.11$ & $\omsig = 0.2$ \\ \hline
$\Omega$ & 0.55 & 1.01 \\ \hline
\sigz\ [mb] & 78.6 & 72.6\\ \hline
$\sigma_{\mathrm{tot}}$ [mb] & 86 & 94.8 \\ \hline
 $\lambda$ & 0.82 & 0.74  \\
\end{tabular}
}
\caption{Parameters used in the parameterisation of the GGCF $P(\sigma_\mathrm{tot})$ distribution.}
\label{tbl:CFParam}
\end{table}
The values for $\lambda$ are chosen to produce  
the above-quoted nucleon--nucleon inelastic cross-section of 70~mb. 
The GGCF $P_{\text{H}}(\sigNN)$\
distributions are shown in the inset of Fig.~\ref{fig:GG}, while the 
resulting $P(\Npart)$ distributions are shown in the main panel of the
figure. 

\label{sec:connection}

To connect an experimental measurement of collision
centrality such as \sumETPb\ to the results of the Glauber or
GGCF Monte Carlo simulation, a model for the \Npart\ dependence of
the \sumETPb\ distribution is required. 
The usual basis for models
previously applied to \nucnuc\ and \pdA\ collisions is the WN
model \cite{Bialas:1976ed}, which predicts that
the average \sumETPb\ increases proportionally to \Npart\ with the
proportionality constant equal to one half the corresponding average
FCal \sumet\ in \pp\ collisions. 

Under the WN model, the \sumETPb\ distribution for fixed \Npart\ would
be obtained from a \Npart-fold convolution of the corresponding
distribution in \pp\ collisions. This
convolution is straightforward if the \sumET\ distribution in \pp\ collisions
is described by a gamma distribution \cite{Tannenbaum:2004xi} 
\begin{equation}
{\rm gamma}\left({\textstyle \sum} E_{\mathrm{T}}; k, \theta\right)
= \frac{1}{\Gamma(k)}\frac{1}{\theta} \left(\frac{\sumET}{\theta}\right)^{k
-1} e^{-\sumET/\theta},
\label{eq:gamma}
\end{equation}
since gamma distributions have the property that an $N$-fold convolution
of a gamma distribution with parameters $k$ and $\theta$ yields another
gamma distribution with the same $\theta$ and a modified $k$
parameter, $k' = Nk$.  

Attempts to fit the measured \sumETPb\ distribution using 
pure WN-convolved gamma distributions and the Glauber \Npart\
distribution yield unphysical results for the nucleon--nucleon
parameters, $k_0$ and $\theta_0$, when those parameters are free
parameters of the fit. In particular, $k_0$ is less than unity, which
implies a \sumET\ distribution that increases with decreasing \sumET\
faster than $e^{-\sumET/\theta_0}$, and $\theta_0$ is unrealistically
large. The resulting nucleon--nucleon \sumET\ distribution is also inconsistent with
that measured in \pp\ collisions \cite{Aad:2012mfa}. The poor behaviour
of the WN model is primarily due to the difference in shape between
the Glauber \Npart\ distribution and the measured \sumETPb\
distribution. To improve the description of the measured \sumETPb\
distribution, a generalisation of the WN model is implemented that
parameterises the \Npart\ dependence of the $k$ and $\theta$
parameters of the gamma distribution as  
\begin{eqnarray}
\knp =  k_0 + k_1 \left(\Npart - 2 \right),\nonumber\\
\thnp =  \theta_0 + \theta_1 \log{\left(\Npart - 1 \right)}\,.
\label{eq:model}
\end{eqnarray}
For $k_1 = k_0/2$ and $\theta_1 = 0$, this model reduces to the WN
model. The $\log\left(\Npart - 1\right)$ term allows for a possible variation in
the effective acceptance of the FCal due to an \Npart-dependent
backward shift in the \pPb\ centre-of-mass system \cite{Steinberg:2007fg}. 

To limit the number of free parameters when fitting the \sumETPb\
distribution, $k_0$ and $\theta_0$ are obtained by fitting the
detector-level \sumETA\ distributions in {\sc Pythia6} and {\sc Pythia8} $pp$ simulations. These
simulations have been shown to give a reasonable 
description of the corresponding \pp\ collision data at $\sqs = 
7$~\TeV\ \cite{Aad:2012mfa} although they both slightly under-predict the
average forward transverse energy. The contribution of electronic noise 
to the simulated distribution was determined by examining the \sumETPb\
distribution in empty beam bunch crossings in data. In the fit, the
gamma distributions were convolved with the effects of this
noise before comparison with the data.
The {\sc Pythia8} fit results, $k_0 = 1.40$ and $\theta_0 = 3.41$, are used
for the default analysis. The {\sc Pythia6} fit results, $k_0 = 1.23$ and
$\theta_0 = 2.68$, are used to evaluate systematic
uncertainties. 

\begin{figure}[!htb]
\centerline{
\includegraphics[width=0.60\textwidth]{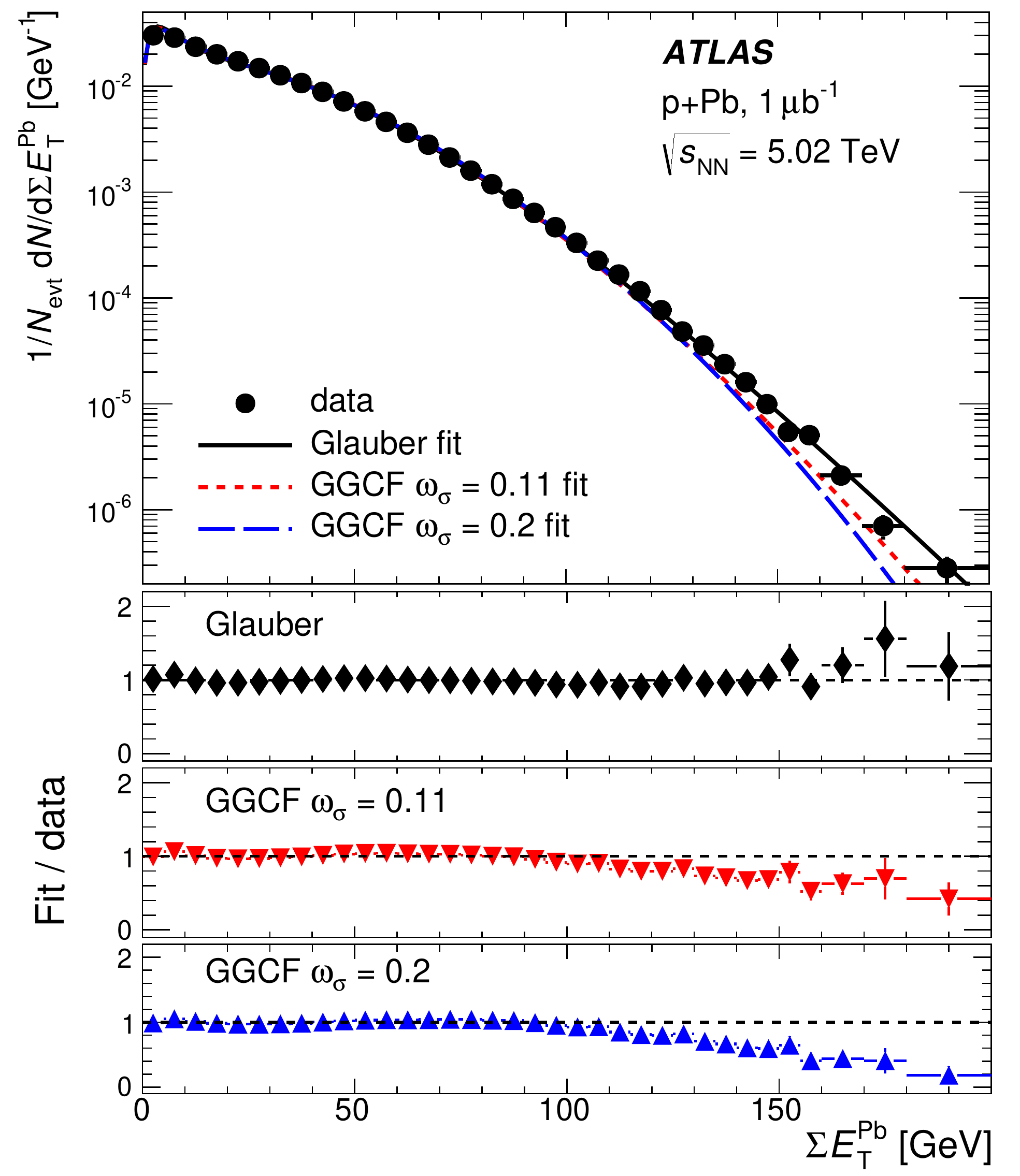}
}
\caption{Top panel: Measured \sumETPb\ distribution compared to Glauber (solid), GGCF with $\omsig = 0.2$ (long dashed), and GGCF with $\omsig = 0.11$ (short dashed) fits. Lower panels: ratios of Glauber and GGCF fit distributions to the data distribution.}
\label{fig:GlauberFits}
\end{figure}

The measured \sumETPb\ distribution is fitted with a distribution
produced by summing the \Npart-dependent gamma distributions, after
weighting them by \Pnpart\ and including an additional convolution to account
for electronic noise. The model distribution is also re-weighted to
properly describe the \sumETPb-dependent event selection efficiency in
the data, which is estimated 
using the {\sc Pythia} MC samples under the assumption that the \pPb\ inefficiency
for a given \sumETPb\ is the same as that in \pp\ collisions. 
Results are shown in Fig.~\ref{fig:GlauberFits} 
for the Glauber model and the two GGCF models.
The fits provide a
good description of the \sumETPb\ distribution for $\sumETPb <
100$~\GeV\ for all three geometric models, although at
higher \sumETPb\ the Glauber fit describes the data better.
The deviations of the GGCF fits from the data
become significant near $\sumETPb = 120$~\GeV; the fraction of the
total \sumETPb\ distribution above this value is approximately 0.1\%.

The parameters $k_1$ and $\theta_1$ are obtained from fixing $k_0$ and
$\theta_0$ and fitting the \sumET distribution to the data. They are
presented in Table~\ref{tbl:fitparams} along with the ratios $k_1/k_0$
for each of the geometric models.
Pure WN behaviour would correspond to $k_1/k_0 = 0.5$ and $\theta_1 =
0$. The results indicate
substantial deviations from WN behaviour for the Glauber and
GGCF $\omsig = 0.2$ fits, while the GGCF $\omsig =
0.11$ fit yields both a $k_1/k_0$ that is close to 0.5 and a small
$\theta_1$. 
The success of the above-described fitting procedure in describing the
measured \sumETPb\ distributions using parameterisations of the \Npart\
dependence of the \sumETPb\ response with only two free parameters is
due to the similarity of the shapes of the $P(\Npart)$ distribution and the
measured \sumETPb\ distributions. However, the pronounced knee in the 
Glauber $P(\Npart)$ distribution requires more non-linearity in
the \Npart\ dependence of the \sumETPb\ response. In contrast, 
the lack of such a feature in the GGCF $P(\Npart)$ distributions allows a
simpler description of the measured \sumETPb\ distribution. 

The results of the fit procedure described above provide a data-driven
estimate of the total \pPb\ event-selection efficiency. For the
default {\sc Pythia8}-based results, the integral of the
simulated \sumETPb\ distribution is 2\% higher than that in the data. 
The deficit in the data is concentrated at low \sumETPb\ values,
consistent with losses due to event selection. However, a
detailed analysis of the residual differences between the best fit
and measured distributions indicates an excess of very low \sumETPb\
events in the data, which varies from $\sim 0\%$ for the Glauber fit to
1.8\% for the GGCF fit with $\omsig = 0.2$. These may arise from
residual diffractive or photo-nuclear collisions and are considered
background. For the purpose of defining \sumETPb\ centrality
intervals, this background effectively increases the event-selection
efficiency by adding events that are all in the 90--100\% centrality
interval. For {\sc Pythia6}, the \sumETPb\ fits using the default model
yield a total efficiency of 97\% and up to 1\% background. The
alternative models for \knp\ and \thnp\ yield a similar total efficiency
of 98\% and a background rate as high as 2\%, a rate that is
compatible with independent estimates of the rate for collisions
involving diffractive excitation of the proton to pass the applied
event selections. Based on these results, the total effective
efficiency including background is then taken to be 98\% and the
uncertainty is conservatively estimated to be 2\%.

\label{sec:mod_results}

The \avgNpart\ values are obtained for each of the centrality
intervals using the results of the fits to the \sumETPb\
distributions. The \avgNpart\ results along with the total systematic
uncertainties, which are described below, are shown in Fig.~\ref{fig:fitnpart} and listed in
Table~\ref{tbl:npartvalues}.

To obtain systematic errors on \Npart\ in
each centrality interval, the maximum positive and negative fractional
variation in \avgNpart\ away from the default results is determined
for different classes of variations, detailed below.

To evaluate the impact of the total event selection uncertainty, new
centrality intervals are chosen assuming a total efficiency of 100\%
and 96\% and the complete analysis is repeated. To account for
possible inaccuracies in the {\sc Pythia8}-simulated $\mathrm{d}\eT/\mathrm{d}\eta$ in the
region of the FCal acceptance, the analysis is repeated separately
under $\pm 10\%$ re-scalings of {\sc Pythia8} \sumETPb\ values commensurate
with the scale of the data--{\sc Pythia8} differences observed in
Ref.~\cite{Aad:2012mfa}. Other variations for which the complete
analysis is repeated are: (i) using the {\sc Pythia6} event generator to fix
$k_0$ and $\theta_0$, (ii) alternative models for \knp\ and \thnp,
(iii) $\pm 5$~mb changes in \sigNN, and (iv) variations in the
parameters of the nuclear density distribution.  For the model
uncertainty, two alternative parameterisations for \knp\ and \thnp\
are used. One of these kept $\theta$ constant, $\thnp = \theta_0$
while allowing for a quadratic dependence of $k$ on \Npart. The other
included both a quadratic term in \knp\ and the logarithmic term
in \thnp\ but fixed $k_1 = k_0/2$ to reduce the number of free
parameters.

The resulting maximal variations are then summed in quadrature over
the different classes separately for the positive and negative
variations to produce the uncertainties listed in
Table~\ref{tbl:npartvalues}. The uncertainties for most centrality intervals
are dominated by comparable contributions from the choice of model
parameterisation, differences between the {\sc Pythia6} and {\sc Pythia8} \sumET\
fits, and uncertainties in \sigNN. For the 40--60\% and 60--90\%
interval the uncertainty in event selection efficiency has a
contribution to the \Npart\ systematic uncertainty that is similar in
magnitude to these other three sources.

\begin{table}
\begin{center}
\begin{tabular}{l|c|c|c}
Glauber model              & $k_1$     & $k_1/k_0$ & $\theta_1$  \\ \hline 
Standard Glauber          & $0.425(2)$  & $0.304(1)$   & $+1.32(1)$   \\ 
GGCF $\omsig = 0.11$ & $0.901(3)$  & $0.643(2)$   & $+0.074(4)$  \\ 
GGCF $\omsig = 0.2$   & $1.139(3)$  & $0.813(2)$   & $-0.209(2)$  \\ 
\end{tabular}
\end{center}
\caption{ Optimal fit parameters and uncertainties obtained from fits to the measured \sumETPb\ distribution. Uncertainties on the fit parameters are shown in parenthesis.}
\label{tbl:fitparams}
\end{table}

\begin{table}
\begin{center}
{\setlength{\extrarowheight}{5pt}
\begin{tabular}{c|c|c|c}
Centrality & Glauber & GGCF $\omsig = 0.11$ & GGCF $\omsig = 0.2$ \\ \hline 
60--90\% & 
$4.0^{+0.2}_{-0.3} \left( ^{+5\%}_{-8\%}\right)$ &
$3.6^{+0.2}_{-0.2} \left( ^{+5\%}_{-5\%}\right)$ &
$3.4^{+0.3}_{-0.2} \left( ^{+8\%}_{-5\%}\right)$ \\
40--60\% &  
$7.4^{+0.4}_{-0.6}     \left( ^{+6\%}_{-8\%}\right)$ & 
$6.6^{+0.4}_{-0.4}     \left( ^{+6\%}_{-6\%}\right)$ & 
$6.3^{+0.5}_{-0.3} \left( ^{+8\%}_{-5\%}\right)$\\
30--40\% &  
$9.8^{+0.6}_{-0.6} \left( ^{+6\%}_{-6\%}\right)$ &
$9.2^{+0.5}_{-0.5} \left( ^{+6\%}_{-6\%}\right)$ &
$8.9^{+0.6}_{-0.5} \left( ^{+7\%}_{-5\%}\right)$\\
20--30\% &  
$11.4^{+0.6}_{-0.6} \left( ^{+6\%}_{-6\%}\right)$ &
$11.2^{+0.6}_{-0.7} \left( ^{+6\%}_{-6\%}\right)$ &
$11.1^{+0.7}_{-0.6} \left( ^{+6\%}_{-6\%}\right)$ \\
10--20\% &  
$13.0^{+0.8}_{-0.7} \left( ^{+6\%}_{-6\%}\right)$ &
$13.7^{+0.8}_{-0.8} \left( ^{+6\%}_{-7\%}\right)$ &
$14.1^{+0.9}_{-0.8} \left( ^{+6\%}_{-6\%}\right)$ \\
5--10\% &  
$14.6^{+1.2}_{-0.8} \left( ^{+8\%}_{-6\%}\right)$ &
$16.5^{+1.0}_{-1.0} \left( ^{+6\%}_{-6\%}\right)$ &
$17.4^{+1.1}_{-1.1} \left( ^{+7\%}_{-6\%}\right)$ \\
1--5\% &  
$16.1^{+1.7}_{-0.9} \left( ^{+11\%}_{-6\%}\right)$ &
$19.5^{+1.3}_{-1.3} \left( ^{+7\%}_{-7\%}\right)$ &
$21.4^{+1.4}_{-2.0} \left( ^{+7\%}_{-9\%}\right)$ \\
0--1\% &  
$18.2^{+2.7}_{-1.0} \left( ^{+15\%}_{-5\%}\right)$ &
$24.1^{+1.6}_{-2.0} \left( ^{+7\%}_{-8\%}\right)$ &
$27.4^{+1.6}_{-4.0} \left( ^{+6\%}_{-16\%}\right)$ \\ \hline
0--90\% &  
$8.4^{+0.5}_{-0.4} \left( ^{+6\%}_{-5\%}\right)$ &
$8.5^{+0.5}_{-0.5} \left( ^{+6\%}_{-5\%}\right)$ &
$8.6^{+0.5}_{-0.4} \left( ^{+6\%}_{-5\%}\right)$  
\end{tabular}
\caption{ \avgNpart\ values for centrality intervals used in this analysis together with asymmetric systematic uncertainties  shown as absolute as well as relative uncertainties.}
\label{tbl:npartvalues}
}
\end{center}
\end{table}

%% file: atlas_authlist.tex
\begin{flushleft}
{\Large The ATLAS Collaboration}

\bigskip

G.~Aad$^{\rm 84}$,
T.~Abajyan$^{\rm 21}$,
B.~Abbott$^{\rm 112}$,
J.~Abdallah$^{\rm 152}$,
S.~Abdel~Khalek$^{\rm 116}$,
O.~Abdinov$^{\rm 11}$,
R.~Aben$^{\rm 106}$,
B.~Abi$^{\rm 113}$,
M.~Abolins$^{\rm 89}$,
O.S.~AbouZeid$^{\rm 159}$
% ...
\bigskip
\\
% ...
$^{11}$ Institute of Physics, Azerbaijan Academy of Sciences, Baku, Azerbaijan\\
% ...
$^{21}$ Physikalisches Institut, University of Bonn, Bonn, Germany\\
% ...
$^{84}$ CPPM, Aix-Marseille Universit{\'e} and CNRS/IN2P3, Marseille, France\\
% ...
$^{89}$ Department of Physics and Astronomy, Michigan State University, East Lansing MI, United States of America\\
% ...
$^{106}$ Nikhef National Institute for Subatomic Physics and University of Amsterdam, Amsterdam, Netherlands\\
% ...
$^{112}$ Homer L. Dodge Department of Physics and Astronomy, University of Oklahoma, Norman OK, United States of America\\
$^{113}$ Department of Physics, Oklahoma State University, Stillwater OK, United States of America\\
% ...
$^{116}$ LAL, Universit{\'e} Paris-Sud and CNRS/IN2P3, Orsay, France\\
% ...
$^{152}$ Institute of Physics, Academia Sinica, Taipei, Taiwan\\
% ...
$^{159}$ Department of Physics, University of Toronto, Toronto ON, Canada\\
% ...
\end{flushleft}
